\documentclass[reprint,
 amsmath,amssymb,
 aps,nofootinbib
]{revtex4-1}
\usepackage{tabularx}
\usepackage{xcolor}
\usepackage[
  margin=2.5cm,
  includefoot,
  footskip=20pt,
]{geometry}
\usepackage[ colorlinks = true,
             linkcolor = blue,
             urlcolor  = blue,
             citecolor = blue,
             anchorcolor = green,
]{hyperref}

\usepackage{graphicx}
\usepackage{dcolumn}
\usepackage{bm}
\usepackage{physics}
\begin{document}

\title{Few-cycle vacuum squeezing in nanophotonics}
\author
 {Rajveer Nehra,$^{1\ast}$, 
  Ryoto Sekine$^{1,\ast}$, Luis Ledezma$^{1,2}$, \\
  Qiushi Guo$^{1}$,
  Robert M. Gray$^{1}$,
  Arkadev Roy$^{1}$, and Alireza Marandi$^{1,\dagger}$
 {{$^1$Department of Electrical Engineering,}\\
 {California Institute of Technology, 
 Pasadena, California 91125, USA.,}}\\
 {{$^2$Jet Propulsion Laboratory},\\
 {{California Institute of Technology, Pasadena, California 91109, USA.}}}
 \\
 {{$^\ast$These authors contributed equally to this work.}}\\
 {{$^\dagger$E-mail: marandi@caltech.edu, rnehra@caltech.edu}}
}
\date{\today}
\begin{abstract}
One of the most fundamental quantum states of light is squeezed vacuum, in which noise in one of the quadratures is less than the standard quantum noise limit. Significant progress has been made in the generation of optical squeezed vacuum and its utilization for numerous applications. However, it remains challenging to generate, manipulate, and measure such quantum states in nanophotonics with performances required for a wide range of scalable quantum information systems. Here, we overcome this challenge in lithium niobate nanophotonics by utilizing ultrashort-pulse phase-sensitive amplifiers for both generation and all-optical measurement of squeezed states on the same chip. 
We generate a squeezed state spanning over more than 25 THz of bandwidth supporting only a few optical cycles, and measure a maximum of 4.9 dB of squeezing ($\sim$11 dB inferred). 
This level of squeezing surpasses the requirements for a wide range of quantum information systems. Our results on generation and measurement of  few-optical-cycle squeezed states in nanophotonics enable a practical path towards scalable quantum information systems with THz clock rates and open opportunities 
for studying non-classical nature of light in the sub-cycle regime. 
\end{abstract}
\maketitle
Quantum information processing offers great promises for computation, secure communication, metrology and sensing. Many physical platforms such as nuclear spins, superconducting circuits, photonics, trapped ions, quantum dots, and neutral atoms have widely been explored in the pursuit to build 
quantum information processors~\cite{clarke2008superconducting,o2009photonic,cirac1995quantum,quantumdots, saffman2010quantum,vandersypen2001experimental}.
Among these, photonics stands out because of its potential for scalability, room-temperature logical operations, and ease of encoding~\cite{Chen2014,roslund2014wavelength, Asavanant2019_alt,Larsen2019_alt,zhong2020quantum}
quantum information in both discrete~\cite{Knill:2001aa} and continuous variables~\cite{lloyd1999quantum,PhysRevA.79.062318}. 

In continuous-variable (CV) quantum photonics, information is encoded in continuous amplitude and phase values of the quantized electromagnetic field. The single-mode and multimode squeezed states are widely used for various applications including quantum-enhanced interferometry such as in LIGO~\cite{yu2020quantum},  microscopy~\cite{casacio2021quantum}, and quantum teleportation~\cite{furusawa1998unconditional}. Moreover, highly entangled CV quantum states, i.e., cluster states~\cite{Asavanant2019_alt, Larsen2019_alt, Chen2014}, serve as a universal resource for one-way quantum computation~\cite{Raussendorf}.

Typically, such high-quality CV states are generated from single or two-mode squeezed vacuum generated using quadratic ($\chi^{(2)}$) parametric processes either in bulk crystals or waveguides with large ($\sim$100 $\mu \text{m}^2$) mode areas~\cite{Chen2014,kanter2002squeezing,roslund2014wavelength,Asavanant2019_alt, larsen2020deterministic}. While such experiments using bulky discrete components have been successful to demonstrate small and medium-scale quantum circuits, it is highly desired to achieve CV quantum states with comparable qualities in nanophotonics to enable large-scale integrated quantum circuits. 

In nanophotonics, silicon nitride (SiN) and silica platforms have been used for many quantum photonic experiments such as entangled photon-pair generation, squeezing, error correction, and small scale Gaussian boson sampling~\cite{joshi2020frequency,vaidya2020broadband,yang2021squeezed, vigliar2021error,arrazola2021quantum}. However, their inherently weak cubic ($\chi^{(3)}$) nonlinearity typically necessitates using high-Q resonators, which imposes limitations on accessible squeezing levels and bandwidths. Despite significant advances, the measured squeezing levels have so far remained below 2 dB in nanophotonics~\cite{dutt2015chip,vaidya2020broadband,yang2021squeezed,cernansky2020nanophotonic}. 

\begin{figure*}
\centering
\includegraphics[width=\linewidth]{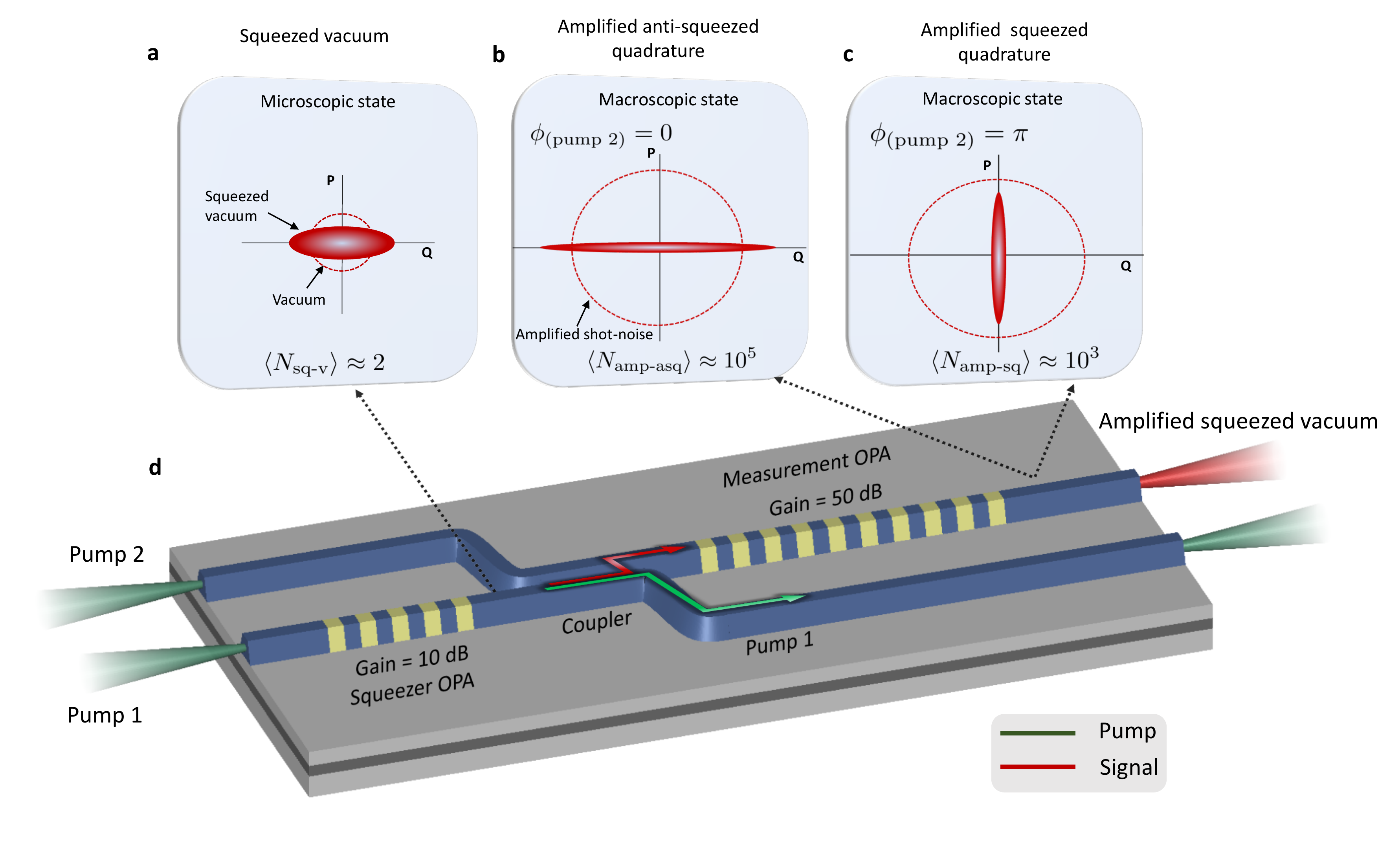}
\caption{\textbf{Illustration of generating squeezed vacuum and measuring it all optically in nanophotonics.} 
Phase-space distributions of \textbf{(a)} phase-quadrature squeezed vacuum, and its amplification in the \textbf{(b)} anti-squeezed and \textbf{(c)} squeezed quadratures. The small (large) dashed circles correspond to shot-noise (amplified shot-noise level, see text) and the small (large) filled ellipses are for squeezed vacuum (amplified squeezed vacuum). \textbf{d}) Sketch of our nanophotonic device consisting of the Squeezer OPA, tapered adiabatic coupler, and Measurement OPA. When pumped, the Squeezer OPA, generates a squeezed vacuum state, which is selectively coupled into an adjacent waveguide. There, it is subsequently amplified by the Measurement OPA to macroscopic power levels.}\label{fig:Fig1}
\end{figure*}

On the other hand, the measurements in CV quantum systems have typically relied on balanced homodyne detection (BHD) using highly-efficient and low-noise photodetectors which are limited in bandwidths to MHz and GHz ranges~\cite{tasker2021silicon}. Moreover, in nanophotonics, the loss associated with transferring the microscopic quantum states from a tightly confined mode to a photodetector has imposed barriers in the measurement capabilities of such states~\cite{Shayan_PRL,PhysRevLett.127.183601,chen2021ultra,vaidya2020broadband}. A potential solution for these measurement challenges lies in all-optical measurement schemes based on a noiseless phase-sensitive amplifier with sufficiently large gain~\cite{caves1982quantum, shaked2018lifting,takanashi2020all,frascella2021overcoming, PhysRevA.101.053801} that can eliminate the limitations of homodyne detection and the sensitivity to detection losses. However, achieving such large gains ($>$30 dB) over broad optical bandwidths is challenging in nanophotonics with cubic nonlinearity~\cite{ye2021overcoming}. 

Recently, lithium niobate (LN) nanophotonics has opened promising avenues in optical communication, sensing, and computation due to its extraordinary optical, electrical, and acoustic properties~\cite{zhu2021integrated, jankowski2021dispersion}. A combination of sub-wavelength confinement of the optical mode, strong $\chi^{(2)}$ nonlinearity, high-fidelity quasi-phasematching (QPM) by periodic poling, and dispersion engineering for longer interaction lengths has enabled devices outperforming the traditional LN devices~\cite{ledezma2021intense, Jankowski:20,  guo2021femtojoule, hu2021chip}.

In this work, we use a nanophotonic circuit in LN and experimentally demonstrate record-level generation and all-optical measurement of ultra-short-pulse squeezed vacuum as the building block of a scalable CV quantum nanophotonics. Our circuit combines two dispersion-engineered phase-sensitive optical parametric amplifiers (OPAs)~\cite{ledezma2021intense}, as shown in Fig.~\ref{fig:Fig1}. The first OPA generates a microscopic squeezed vacuum which is then amplified with a high-gain OPA to macroscopic levels within the same nanophotonic chip. The resulting macroscopic field carries information about the microscopic squeezed state, which can be measured with a high tolerance to loss.

In Figure~\ref{fig:Fig1}a, the phase-space distributions for vacuum (dashed circle) and phase-quadrature squeezed vacuum (filled ellipse) are displayed. For the figure, we consider a particular case of 10-dB squeezing with mean photon-number $\langle N_{\text{sq-v}} \rangle \approx 2$. The squeezed light is then amplified by an OPA with 50 dB of phase-sensitive gain, thereby amplifying the few-photon squeezed signal to a macroscopic power level. In Fig.~\ref{fig:Fig1}b and 1c, the phase-space distributions corresponding to amplified anti-squeezed and amplified squeezed quadratures  are shown for two particular pump phases of the Measurement OPA, $\phi_{\text{ Pump 2}}$. 
The dashed circles in Fig.~\ref{fig:Fig1}b and 1c represent the amplified shot-noise level corresponding to the phase-space distribution of amplified vacuum. For brevity, we represent the distribution of the amplified vacuum as a circle because the vacuum state is phase-insensitive and therefore, the amplification of any phase space quadrature by the Measurement OPA will lead to the same noise level.
In the case of Fig.~\ref{fig:Fig1}b with $\phi_{\text{ Pump 2}}=0$, the anti-squeezed quadrature (Q) is amplified while the orthogonal phase quadrature (P) is de-amplified such that the output field is dominated entirely by the Q quadrature and the P quadrature can be considered negligible. In such a high-gain amplification regime, the total average photon-number (power) of the output field is: $\langle \hat{N}_+\rangle \propto \langle \hat{Q}_{\text{amp}}^2 \rangle \approx \mathcal{O}(10^5)$. Likewise, by changing the pump phase of the Measurement OPA to $\phi_{\text{ Pump 2}}=\pi$, the original squeezed quadrature is amplified to achieve $\langle \hat{N}_- \rangle \propto \langle \hat{P}_{\text{amp}}^2 \rangle \approx \mathcal{O}(10^3)$, as shown in Figure~\ref{fig:Fig1}c. As a result, the macroscopic output of the Measurement OPA provides a direct all-optical measurement of the microscopic squeezed state. The squeezing ($S_-$) and anti-squeezing ($S_+$) can then be determined as  $S_\pm[\text{dB}] = 10\text{log}_{10}[\langle \hat{N}_{\pm}\rangle /\langle \hat{N}_{\text{v}}\rangle ]$, where $\langle \hat{N}_{\text{v}}\rangle $ denotes the amplified vacuum. In the ideal case, the squeezing
(anti-squeezing) can be determined as~\cite{SuppMat}
\begin{equation}
    S_\pm[\text{dB}] = 10\text{log}\bigg(\frac{\text{sinh}^2(r_2\pm r_1)}{\text{sinh}^2r_2}\bigg), 
\end{equation}
\label{eq:ideal_sq}
where $r_1$ and $r_2$ are the gain parameters for the Squeezer and Measurement OPAs, respectively. Sufficient gain ($>$ 33 dB for $\sim$ 11 dB of Squeezer OPA gain, see Ref.~\cite{SuppMat}, Part 6 for details) in the Measurement OPA allows a direct measurement of the phase-squeezed vacuum generated in the Squeezer OPA~\cite{SuppMat}.
Importantly, the high-gain Measurement OPA makes our measurement tolerant to coupling losses and photodetection inefficiencies as high as $\sim$ 7 dB~\cite{SuppMat}.

In experiments, the Squeezer (low-gain) and Measurement (high-gain) OPAs of our circuit are periodically-poled with the lengths of 2.5 mm and 5.0 mm, respectively. The output of the Squeezer OPA (microscopic squeezed vacuum) is coupled to the Measurement OPA through a directional coupler. To make our directional coupler broadband and less susceptible to fabrication imperfections, we employ an adiabatic design where both of the waveguides are tapered while keeping the gap constant throughout the coupler length. The coupler directs the squeezed vacuum to the adjacent waveguide towards the Measurement OPA, and keeps the residual pump of the Squeezer OPA in the original waveguide as shown in Fig.~\ref{fig:Fig1}d. In our current device, the coupler causes $\sim$30\%  loss for the squeezed vacuum, and leaks $\sim$20\% of the squeezer pump to the Measurement OPA~\cite{SuppMat}. Our numerical simulations suggest that the coupling performance can be significantly improved to $>95\%$ for squeezed signal and $<5\%$ for squeezer pump, which will lead to better measurement quality~\cite{SuppMat}. 

\begin{figure*}
\centering
\includegraphics[width = 0.78\textwidth]{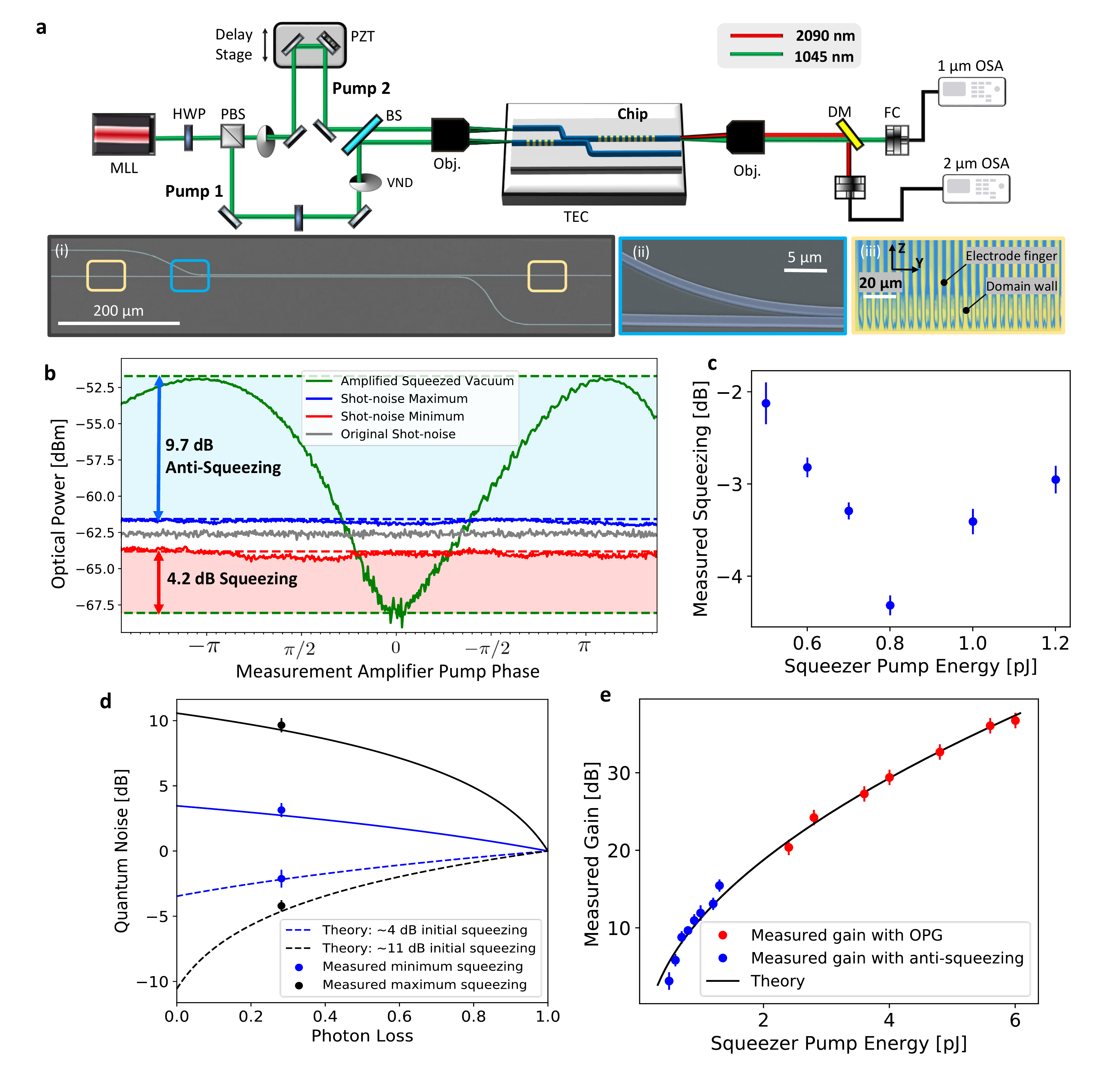}
\caption{\textbf{Generation and measurement of squeezed light in the LN nanophotonic chip.} 
\textbf{a}) Experimental setup. A MLL is split to pump the squeezer (Pump 1) and measurement (Pump 2) OPAs. The phase difference between the pumps is modulated by a piezoelectric transducer (PZT) in the Pump 2 arm. At the output of the nanophotonic chip, light from the measurement amplifier waveguide is collected. The amplified squeezed signal (red) and residual Pump 2 (green) are separated using a DM, and measured by two different OSAs. (i) a false-colored SEM image of our nanophotonic circuit, (ii) a zoomed-in SEM image of the coupler region, and (iii) a false-colored second harmonic microscope image of the periodically-poled region prior to etching the waveguides. 
 \textbf{b}) Squeezing measurement by an OSA in the zero-span mode at 2090 nm. The green trace shows the amplified squeezed vacuum when the PZT is modulated by a ramp signal. The other traces represent amplified vacuum (shot-noise) when Pump 1 is blocked at different Pump 2 powers (see main text). \textbf{c}) The squeezing measured at 2090 nm  for several values of Pump 1 while keeping Pump 2 constant. \textbf{d}) Loss analysis of the squeezing measurements. The solid (dashed) curves show the degradation of anti-squeezing (squeezing) as the photon loss increases and the solid data points correspond to measured values  of minimum and maximum squeezing. \textbf{e}) The squeezer gain dependence on the energy of Pump 1. Blue points are measured from anti-squeezing and red points are directly obtained from optical parametric generation (OPG) measurements. Error bars are obtained from the statistics of the measurements.
 MLL: mode-locked laser, PBS: polarized beamsplitter, BS: beam splitter, PZT: piezoelelectric transducer, HWP: half-wave plate, DM: dichroic mirror, Obj.: reflective objective, VND: variable neutral-density filter, FC: fiber coupler, OSA: optical spectrum analyzer, TEC: thermoelectric cooler.}
 \label{fig:Fig2}
\end{figure*}

Our experimental setup is shown in Fig.~\ref{fig:Fig2}a. The Squeezer and Measurement OPAs are pumped by a mode-locked laser (Menlo Systems Orange A) generating $\sim$75-fs-long nearly transform-limited pulses at a 250-MHz repetition rate. The relative phase between Pump 1 (Squeezer OPA) and Pump 2 (Measurement OPA) pulses is modulated by a piezoelectric transducer (PZT) on the Pump 2 arm.  At the output of the nanophotonic chip, the amplified squeezed signal and Measurement OPA pump are first separated using a dichroic mirror and then are detected by two different optical spectrum analysers (OSAs). In Fig.~\ref{fig:Fig2}a, we show, (i) a false-colored scanning electron microscope (SEM) image of our nanophotonic circuit, (ii) a zoomed-in SEM image of the coupler region, and (iii) a false-colored second harmonic microscope image of the periodically-poled region prior to etching the waveguides. 


Figure~\ref{fig:Fig2}b shows an example measurement of our squeezed state. The green trace shows the output signal of the Measurement OPA using an OSA in a zero-span mode at 2090 nm while keeping both Pump 1 and Pump 2 on and modulating the PZT by a 1-Hz ramp signal. To accurately measure the squeezing, we need to eliminate the effect of residual interference of the two pumps at the output of the measurement. We achieve this by determining the maximum and minimum of this residual interference and then calibrating our amplified shot-noise levels by subsequently varying the power of Pump 2 to these maximum and minimum pump powers while blocking Pump 1. These two levels of Pump 2 result in ``shot-noise maximum'' and ``shot-noise minimum'' as shown in Fig.~\ref{fig:Fig2}b, while ``original shot-noise'' corresponds to the Pump 2 level during the squeezing measurement. Hence, in the squeezing measurement, the shaded area below (above) the ``shot-noise minimum'' (``shot-noise maximum'') corresponds to  squeezing (anti-squeezing) at the input of the high-gain OPA. A detailed discussion on our shot-noise calibration measurements can be found in Ref.~\cite{SuppMat}, Part 2.

We further characterize the dependence of squeezing at 2090 nm on the pump power while keeping the Pump 2 constant and performing the shot-noise calibration for each power level as shown in Fig.~\ref{fig:Fig2}c. As we increase the pump power in the Squeezer OPA, the level of measured squeezing increases at first. However, above 0.8 pJ of pump pulse energy, we observe that further increasing the squeezer pump decreases the level of measured squeezing. The degradation of measured squeezing at high pump powers may be due to the existence of a small phase noise and relative chirp between Pump 1 and Pump 2 which can mix the loss-degraded squeezed quadrature with the relatively large anti-squeezed quadrature~\cite{PhysRevA.52.4202,oelker2016ultra}. Additionally, parasitic nonlinear effects such as the photorefractive effect and nonlinear absorption mechanisms in the waveguide can also account for the degradation of squeezing at higher pump powers.

Figure~\ref{fig:Fig2}d shows how squeezing levels degrade in the presence of photon loss $(1-\eta)$. The solid and dashed curves represent anti-squeezing and squeezing, respectively. Analytically, $S_{\pm}[\text{dB}] = 10\text{log}[(1-\eta) + \eta e^{\pm2r}]$, where $(1-\eta)$ determines the loss present in the detection protocol and $r$ is the squeezing parameter characterizing nonlinear interaction strength~\cite{SuppMat}. The solid dots are the experimental data points for the minimum and maximum amount of measured squeezing at 2090 nm in Fig.~\ref{fig:Fig2}c. From these measurements, we estimate the total loss $L = 1-\eta \approx 0.3$ experienced by the microscopic squeezed signal before being fully amplified by the measurement OPA. This agrees well with our measured coupling efficiency of the adiabatic coupler using an auxiliary signal centered at 2090 nm (See Ref.~\cite{SuppMat}, Part 4). From the fit, we infer that we successfully generated 10.6 dB of squeezing with the pump energy of $<$1 pJ.  This paves the way for fault-tolerant CV quantum processors in LN nanophotonics, as 10.0 dB of squeezing is sufficient for many architectures including recent proposals with Gottesman-Kitaev-Preskill (GKP) qubit encodings ~\cite{bourassa2021blueprint,PhysRevX.8.021054}.

Figure~\ref{fig:Fig2}e depicts the gain in the Squeezer OPA as a function of Pump 1 pulse energy . The gain for lower pump energies ($<$2.4 pJ) is determined from the anti-squeezing measurements, while for higher pump energies ($>$2.4 pJ) we obtain the gain from a direct measurement of average photon-number~\cite{SuppMat,ledezma2021intense}. For a vacuum-seeded phase-sensitive OPA, the average number of photons in the high parametric gain regime ($\langle\hat{N}\rangle \sim G/4$)  allows us to estimate the gain~\cite{SuppMat}. The solid curve is the fit that includes the overall detection efficiency (including off-chip coupling losses and imperfect detection after the Measurement OPA) and the nonlinear strength as fitting parameters. From the fit, we extract the overall detection efficiency of $\eta^{\text{off-chip}}_{\text{overall}}\sim 0.20$~\cite{SuppMat}. This level of linear loss puts an upper limit of $<1$ dB to the measurable squeezing for a standard BHD. Remarkably, this is not a limiting factor for our all-optical squeezing measurements because of the noiseless amplification by the Measurement OPA.  Note that such lossy measurements are even more detrimental for highly squeezed states, as they are extremely sensitive to losses. This can be seen in Fig.~\ref{fig:Fig2}d, where $\sim$11 dB of initial squeezing degrades by $\sim$10 dB in the presence of the detection losses of $L^{\text{off-chip}}_{\text{overall}} = 1-\eta^{\text{off-chip}}_{\text{overall}} = 0.80$. However, our all-optical measurement is not affected by $L^{\text{off-chip}}_{\text{overall}}$  losses due to the amplification by the Measurement OPA and allows us to measure the squeezing levels as high as 4.9 dB~\cite{SuppMat}. Thus, our measured squeezing is mostly limited by the coupling losses at the adiabatic coupler, which can be fabricated with losses of $<0.05$, as suggested by our numerical simulations~\cite{SuppMat}.

\begin{figure*}
\centering
\includegraphics[width=1\linewidth]{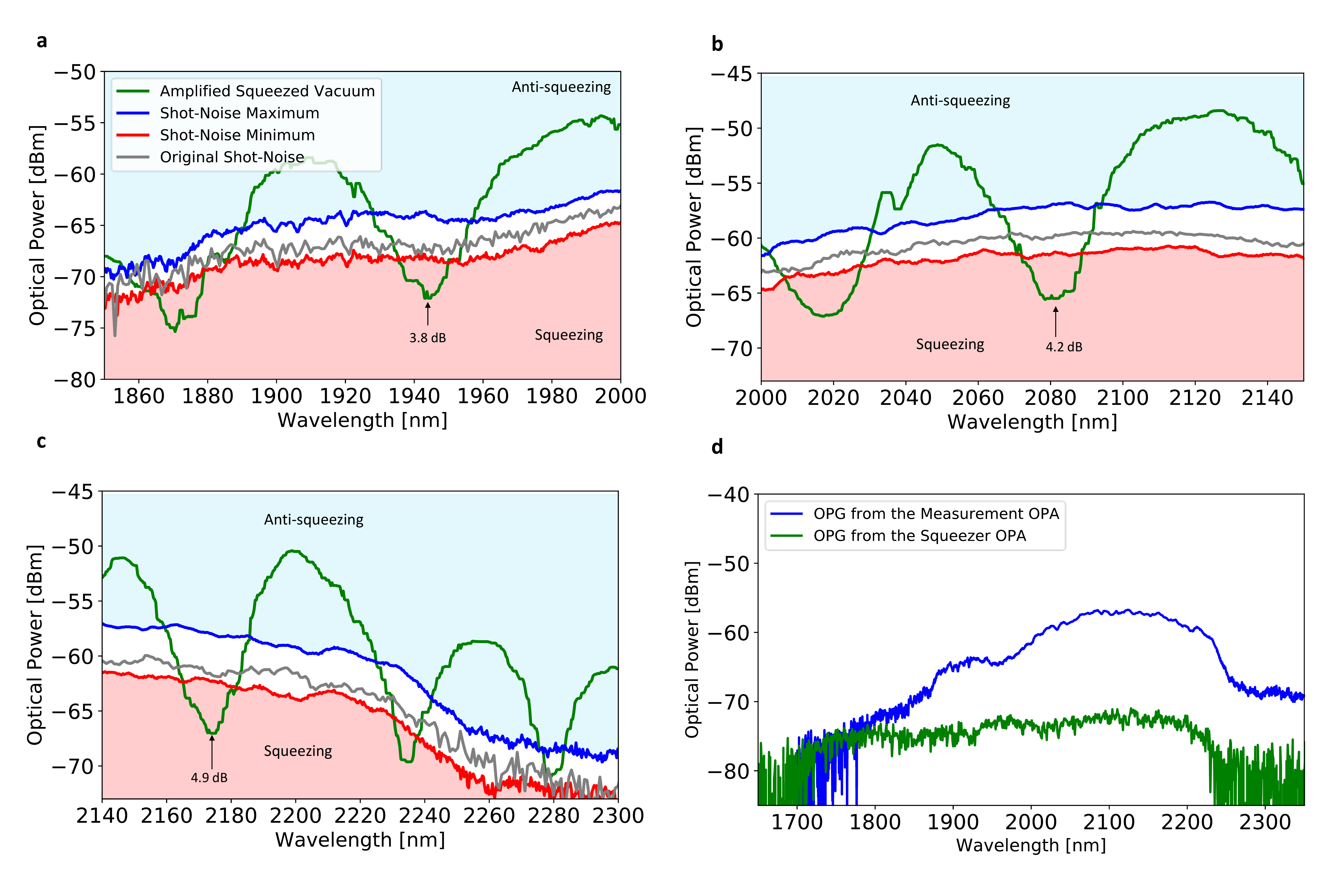}
\caption{\textbf{Broadband squeezing measurements}. \textbf{a, b, c}) Three measurements over different selected bandwidths of the OSA when the PZT is modulated with a 300 mHz ramp signal. The shot noise traces (blue, black and red) were acquired with Pump 1 blocked. \textbf{d}) Optical parametric generation from the squeezer OPA (green) and measurement amplifier OPA (blue). Both traces are acquired at $\sim$ 6 pJ of pulse energy.}\label{fig:fig3}
\end{figure*}
Figures~\ref{fig:fig3}a, 3b, and 3c show the measured squeezing over a broad bandwidth. The amplified shot-noise is calibrated using the same method as discussed earlier over the entire spectrum. Green traces correspond to measurements by the OSA over three different spectral windows when the PZT is modulated by a slow ramp signal at 300 mHz. Squeezing is present over the entire spectrum with a slight spectral dependence. The measured squeezing is 3.8 dB around 1950 nm, 4.2 dB around 2090 nm, and 4.9 dB around 2200 nm. This is attributed to the wavelength dependence of the coupling efficiency of our adiabatic coupler~\cite{SuppMat}. We measured the squeezing bandwidth to be 25.1 THz. 
 The bandwidth is expected to increase to 36.4 THz, 
 as confirmed by the optical parametric generation (OPG) from the Squeezer OPA in Fig.~\ref{fig:fig3}d. The measured squeezing bandwidth is mostly limited by the slight mismatch of Measurement OPA gain in the wings of the spectrum, as evident from its OPG signal. Due to this difference in the gain spectrum,  the Measurement OPA does not amplify the squeezed vacuum over its entire generation bandwidth to macroscopic levels, leading to a reduced measured squeezing bandwidth. These measurements indicate that our generated squeezed state can occupy a record-level time window of $\sim$ 4 optical cycles (See Ref.~\cite{SuppMat}, Part 3), thereby opening many promising opportunities in ultra-high-speed optical quantum information processing.

\begin{figure}
\centering
\includegraphics[width=1\linewidth]{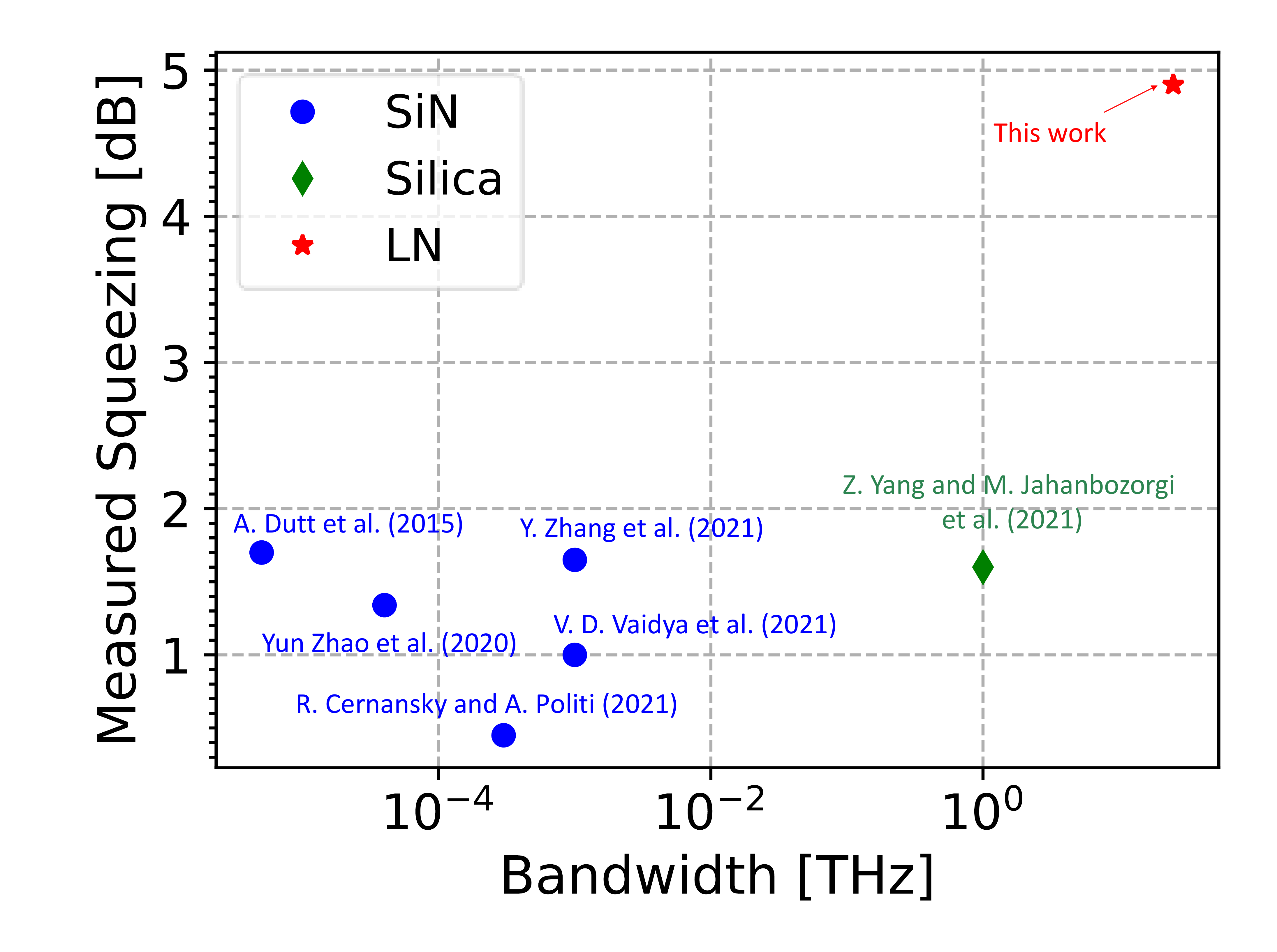}
\caption{\textbf{Performance comparison of measured squeezing levels and squeezing bandwidths on various  nanophotonic platforms.}}
\label{fig:fig5}
\end{figure}

Our results on generating and measuring squeezed states all-optically in nanophotonics mark an important milestone for achieving scalable CV quantum photonic systems. Figure~\ref{fig:fig5} compares our measured squeezing and bandwidth with state-of-the-art demonstrations in nanophotonic platforms including SiN and Silica~\cite{SuppMat}. These experimental demonstrations utilize relatively weaker $\chi^{(3)}$ nonlinearities and require microresonators for enhancement. Additionally, the squeezed light is typically detected using off-chip BHDs which impose limitations on the measured squeezing (due to off-chip coupling losses) and the accessible squeezing bandwidths. Our measured squeezing substantially surpasses these other works both in the magnitude and bandwidth. 

In summary, we demonstrated few-cycle vacuum squeezing in the LN nanophotonic platform. Our results represent a paradigm shift for on-chip squeezed light sources for quantum information processing. The extraordinary performance of quadratic parametric amplifiers is achieved by spatio-temporal confinement of the pump and signal pulses. Immediate improvements would include reducing the coupler loss, improving the input coupling losses (which are currently limiting the gain of the Measurement OPA), and improving the phase noise of the pump.  

All-optical measurements through high-gain amplification brings the microscopic signal to macroscopic levels, making it resilient to coupling losses and imperfect detection, and allows one to measure squeezing over the optical bandwidths that are not accessible through conventional homodyne detection~\cite{Chen2014, yang2021squeezed}. The broadband gain allows for the generation and measurement of squeezing over a bandwidth corresponding to a few-cycle temporal pulse. This can be beneficial for time-multiplexed quantum systems~\cite{Asavanant2019_alt, larsen2020deterministic, marandi2014network} by enabling THz clock rates. Moreover, our results can open paths toward studying quantum fields in the extremely short-pulse regime~\cite{kizmann2019subcycle,riek2017subcycle}. 

Our all-optical measurement in the same nanophotonic circuit eliminates the need for the spatio-temporal shaping of the local oscillator which is typically done in squeezing experiments~\cite{Prem_Kumar_squeezing, kanter2002squeezing}. Such an all-optical measurement can greatly simplify the quadrature measurements for CV quantum information processing using quantum frequency combs ~\cite{Chen2014, roslund2014wavelength}. 
Our all-optical loss-tolerant measurement scheme can also be used to purify noisy quantum measurements for various applications in optical quantum information processing, including all-optical quantum communication, and for protection and characterization of quantum resources~\cite{PhysRevA.82.042315, PhysRevResearch.3.033118, ralph1999all}.

\section*{Data Availability}
The data supporting the plots within this paper and other findings of this study
are available from the corresponding author upon reasonable request.
\section*{Code Availability}
The computer codes used to perform the numerical simulations in this paper are available from the corresponding author upon reasonable request.
\section*{Acknowledgements}
The device nanofabrication was performed at the Kavli Nanoscience Institute (KNI) at Caltech. The authors gratefully acknowledge support from ARO grant no. W911NF-18-1-0285, NSF grant no. 1846273 and 1918549, AFOSR award FA9550-20-1-0040, and NASA/JPL. This project was funded in part by the President’s and Director’s Research and Development Fund of Caltech and JPL. The authors wish to thank NTT Research for their financial and technical support. The authors thank Carlos Gonz\`{a}lez-Arciniegas and Olivier Pfister for fruitful discussions. 
\section*{Authors Contributions}
R.N. and A.M. conceived the idea and designed the experiments; R.N. designed the devices with assistance from L.L. and Q.G.; R.S. fabricated the devices and L.L. performed the periodic poling; R.N. carried out the experiments with assistance from R.S., R.M.G., Q.G. and A.R.; R.N. performed the theoretical and numerical analysis with contributions from L.L.; R.N. and A.M. wrote the manuscript with input from all other authors. A.M. supervised the project.

\bibliography{scibib.bib}

\begin{thebibliography}{55}%
\makeatletter
\providecommand \@ifxundefined [1]{%
 \@ifx{#1\undefined}
}%
\providecommand \@ifnum [1]{%
 \ifnum #1\expandafter \@firstoftwo
 \else \expandafter \@secondoftwo
 \fi
}%
\providecommand \@ifx [1]{%
 \ifx #1\expandafter \@firstoftwo
 \else \expandafter \@secondoftwo
 \fi
}%
\providecommand \natexlab [1]{#1}%
\providecommand \enquote  [1]{``#1''}%
\providecommand \bibnamefont  [1]{#1}%
\providecommand \bibfnamefont [1]{#1}%
\providecommand \citenamefont [1]{#1}%
\providecommand \href@noop [0]{\@secondoftwo}%
\providecommand \href [0]{\begingroup \@sanitize@url \@href}%
\providecommand \@href[1]{\@@startlink{#1}\@@href}%
\providecommand \@@href[1]{\endgroup#1\@@endlink}%
\providecommand \@sanitize@url [0]{\catcode `\\12\catcode `\$12\catcode
  `\&12\catcode `\#12\catcode `\^12\catcode `\_12\catcode `\%12\relax}%
\providecommand \@@startlink[1]{}%
\providecommand \@@endlink[0]{}%
\providecommand \url  [0]{\begingroup\@sanitize@url \@url }%
\providecommand \@url [1]{\endgroup\@href {#1}{\urlprefix }}%
\providecommand \urlprefix  [0]{URL }%
\providecommand \Eprint [0]{\href }%
\providecommand \doibase [0]{http://dx.doi.org/}%
\providecommand \selectlanguage [0]{\@gobble}%
\providecommand \bibinfo  [0]{\@secondoftwo}%
\providecommand \bibfield  [0]{\@secondoftwo}%
\providecommand \translation [1]{[#1]}%
\providecommand \BibitemOpen [0]{}%
\providecommand \bibitemStop [0]{}%
\providecommand \bibitemNoStop [0]{.\EOS\space}%
\providecommand \EOS [0]{\spacefactor3000\relax}%
\providecommand \BibitemShut  [1]{\csname bibitem#1\endcsname}%
\let\auto@bib@innerbib\@empty
\bibitem [{\citenamefont {Clarke}\ and\ \citenamefont
  {Wilhelm}(2008)}]{clarke2008superconducting}%
  \BibitemOpen
  \bibfield  {author} {\bibinfo {author} {\bibfnamefont {J.}~\bibnamefont
  {Clarke}}\ and\ \bibinfo {author} {\bibfnamefont {F.~K.}\ \bibnamefont
  {Wilhelm}},\ }\href@noop {} {\bibfield  {journal} {\bibinfo  {journal}
  {Nature}\ }\textbf {\bibinfo {volume} {453}},\ \bibinfo {pages} {1031}
  (\bibinfo {year} {2008})}\BibitemShut {NoStop}%
\bibitem [{\citenamefont {O'brien}\ \emph {et~al.}(2009)\citenamefont
  {O'brien}, \citenamefont {Furusawa},\ and\ \citenamefont
  {Vu{\v{c}}kovi{\'c}}}]{o2009photonic}%
  \BibitemOpen
  \bibfield  {author} {\bibinfo {author} {\bibfnamefont {J.~L.}\ \bibnamefont
  {O'brien}}, \bibinfo {author} {\bibfnamefont {A.}~\bibnamefont {Furusawa}}, \
  and\ \bibinfo {author} {\bibfnamefont {J.}~\bibnamefont
  {Vu{\v{c}}kovi{\'c}}},\ }\href@noop {} {\bibfield  {journal} {\bibinfo
  {journal} {Nature Photonics}\ }\textbf {\bibinfo {volume} {3}},\ \bibinfo
  {pages} {687} (\bibinfo {year} {2009})}\BibitemShut {NoStop}%
\bibitem [{\citenamefont {Cirac}\ and\ \citenamefont
  {Zoller}(1995)}]{cirac1995quantum}%
  \BibitemOpen
  \bibfield  {author} {\bibinfo {author} {\bibfnamefont {J.~I.}\ \bibnamefont
  {Cirac}}\ and\ \bibinfo {author} {\bibfnamefont {P.}~\bibnamefont {Zoller}},\
  }\href@noop {} {\bibfield  {journal} {\bibinfo  {journal} {Physical review
  letters}\ }\textbf {\bibinfo {volume} {74}},\ \bibinfo {pages} {4091}
  (\bibinfo {year} {1995})}\BibitemShut {NoStop}%
\bibitem [{\citenamefont {Imamog\ifmmode\bar\else\textasciimacron\fi{}lu}\
  \emph {et~al.}(1999)\citenamefont
  {Imamog\ifmmode\bar\else\textasciimacron\fi{}lu}, \citenamefont {Awschalom},
  \citenamefont {Burkard}, \citenamefont {DiVincenzo}, \citenamefont {Loss},
  \citenamefont {Sherwin},\ and\ \citenamefont {Small}}]{quantumdots}%
  \BibitemOpen
  \bibfield  {author} {\bibinfo {author} {\bibfnamefont {A.}~\bibnamefont
  {Imamog\ifmmode\bar\else\textasciimacron\fi{}lu}}, \bibinfo {author}
  {\bibfnamefont {D.~D.}\ \bibnamefont {Awschalom}}, \bibinfo {author}
  {\bibfnamefont {G.}~\bibnamefont {Burkard}}, \bibinfo {author} {\bibfnamefont
  {D.~P.}\ \bibnamefont {DiVincenzo}}, \bibinfo {author} {\bibfnamefont
  {D.}~\bibnamefont {Loss}}, \bibinfo {author} {\bibfnamefont {M.}~\bibnamefont
  {Sherwin}}, \ and\ \bibinfo {author} {\bibfnamefont {A.}~\bibnamefont
  {Small}},\ }\href {\doibase 10.1103/PhysRevLett.83.4204} {\bibfield
  {journal} {\bibinfo  {journal} {Phys. Rev. Lett.}\ }\textbf {\bibinfo
  {volume} {83}},\ \bibinfo {pages} {4204} (\bibinfo {year}
  {1999})}\BibitemShut {NoStop}%
\bibitem [{\citenamefont {Saffman}\ \emph {et~al.}(2010)\citenamefont
  {Saffman}, \citenamefont {Walker},\ and\ \citenamefont
  {M{\o}lmer}}]{saffman2010quantum}%
  \BibitemOpen
  \bibfield  {author} {\bibinfo {author} {\bibfnamefont {M.}~\bibnamefont
  {Saffman}}, \bibinfo {author} {\bibfnamefont {T.~G.}\ \bibnamefont {Walker}},
  \ and\ \bibinfo {author} {\bibfnamefont {K.}~\bibnamefont {M{\o}lmer}},\
  }\href@noop {} {\bibfield  {journal} {\bibinfo  {journal} {Reviews of modern
  physics}\ }\textbf {\bibinfo {volume} {82}},\ \bibinfo {pages} {2313}
  (\bibinfo {year} {2010})}\BibitemShut {NoStop}%
\bibitem [{\citenamefont {Vandersypen}\ \emph {et~al.}(2001)\citenamefont
  {Vandersypen}, \citenamefont {Steffen}, \citenamefont {Breyta}, \citenamefont
  {Yannoni}, \citenamefont {Sherwood},\ and\ \citenamefont
  {Chuang}}]{vandersypen2001experimental}%
  \BibitemOpen
  \bibfield  {author} {\bibinfo {author} {\bibfnamefont {L.~M.}\ \bibnamefont
  {Vandersypen}}, \bibinfo {author} {\bibfnamefont {M.}~\bibnamefont
  {Steffen}}, \bibinfo {author} {\bibfnamefont {G.}~\bibnamefont {Breyta}},
  \bibinfo {author} {\bibfnamefont {C.~S.}\ \bibnamefont {Yannoni}}, \bibinfo
  {author} {\bibfnamefont {M.~H.}\ \bibnamefont {Sherwood}}, \ and\ \bibinfo
  {author} {\bibfnamefont {I.~L.}\ \bibnamefont {Chuang}},\ }\href@noop {}
  {\bibfield  {journal} {\bibinfo  {journal} {Nature}\ }\textbf {\bibinfo
  {volume} {414}},\ \bibinfo {pages} {883} (\bibinfo {year}
  {2001})}\BibitemShut {NoStop}%
\bibitem [{\citenamefont {Chen}\ \emph {et~al.}(2014)\citenamefont {Chen},
  \citenamefont {Menicucci},\ and\ \citenamefont {Pfister}}]{Chen2014}%
  \BibitemOpen
  \bibfield  {author} {\bibinfo {author} {\bibfnamefont {M.}~\bibnamefont
  {Chen}}, \bibinfo {author} {\bibfnamefont {N.~C.}\ \bibnamefont {Menicucci}},
  \ and\ \bibinfo {author} {\bibfnamefont {O.}~\bibnamefont {Pfister}},\ }\href
  {\doibase 10.1103/PhysRevLett.112.120505} {\bibfield  {journal} {\bibinfo
  {journal} {Phys. Rev. Lett.}\ }\textbf {\bibinfo {volume} {112}},\ \bibinfo
  {pages} {120505} (\bibinfo {year} {2014})}\BibitemShut {NoStop}%
\bibitem [{\citenamefont {Roslund}\ \emph {et~al.}(2014)\citenamefont
  {Roslund}, \citenamefont {De~Araujo}, \citenamefont {Jiang}, \citenamefont
  {Fabre},\ and\ \citenamefont {Treps}}]{roslund2014wavelength}%
  \BibitemOpen
  \bibfield  {author} {\bibinfo {author} {\bibfnamefont {J.}~\bibnamefont
  {Roslund}}, \bibinfo {author} {\bibfnamefont {R.~M.}\ \bibnamefont
  {De~Araujo}}, \bibinfo {author} {\bibfnamefont {S.}~\bibnamefont {Jiang}},
  \bibinfo {author} {\bibfnamefont {C.}~\bibnamefont {Fabre}}, \ and\ \bibinfo
  {author} {\bibfnamefont {N.}~\bibnamefont {Treps}},\ }\href@noop {}
  {\bibfield  {journal} {\bibinfo  {journal} {Nature Photonics}\ }\textbf
  {\bibinfo {volume} {8}},\ \bibinfo {pages} {109} (\bibinfo {year}
  {2014})}\BibitemShut {NoStop}%
\bibitem [{\citenamefont {Asavanant}\ \emph {et~al.}(2019)\citenamefont
  {Asavanant}, \citenamefont {Shiozawa}, \citenamefont {Yokoyama},
  \citenamefont {Charoensombutamon}, \citenamefont {Emura}, \citenamefont
  {Alexander}, \citenamefont {Takeda}, \citenamefont {Yoshikawa}, \citenamefont
  {Menicucci}, \citenamefont {Yonezawa} \emph {et~al.}}]{Asavanant2019_alt}%
  \BibitemOpen
  \bibfield  {author} {\bibinfo {author} {\bibfnamefont {W.}~\bibnamefont
  {Asavanant}}, \bibinfo {author} {\bibfnamefont {Y.}~\bibnamefont {Shiozawa}},
  \bibinfo {author} {\bibfnamefont {S.}~\bibnamefont {Yokoyama}}, \bibinfo
  {author} {\bibfnamefont {B.}~\bibnamefont {Charoensombutamon}}, \bibinfo
  {author} {\bibfnamefont {H.}~\bibnamefont {Emura}}, \bibinfo {author}
  {\bibfnamefont {R.~N.}\ \bibnamefont {Alexander}}, \bibinfo {author}
  {\bibfnamefont {S.}~\bibnamefont {Takeda}}, \bibinfo {author} {\bibfnamefont
  {J.-i.}\ \bibnamefont {Yoshikawa}}, \bibinfo {author} {\bibfnamefont {N.~C.}\
  \bibnamefont {Menicucci}}, \bibinfo {author} {\bibfnamefont {H.}~\bibnamefont
  {Yonezawa}},  \emph {et~al.},\ }\href@noop {} {\bibfield  {journal} {\bibinfo
   {journal} {Science}\ }\textbf {\bibinfo {volume} {366}},\ \bibinfo {pages}
  {373} (\bibinfo {year} {2019})}\BibitemShut {NoStop}%
\bibitem [{\citenamefont {Larsen}\ \emph {et~al.}(2019)\citenamefont {Larsen},
  \citenamefont {Guo}, \citenamefont {Breum}, \citenamefont
  {Neergaard-Nielsen},\ and\ \citenamefont {Andersen}}]{Larsen2019_alt}%
  \BibitemOpen
  \bibfield  {author} {\bibinfo {author} {\bibfnamefont {M.~V.}\ \bibnamefont
  {Larsen}}, \bibinfo {author} {\bibfnamefont {X.}~\bibnamefont {Guo}},
  \bibinfo {author} {\bibfnamefont {C.~R.}\ \bibnamefont {Breum}}, \bibinfo
  {author} {\bibfnamefont {J.~S.}\ \bibnamefont {Neergaard-Nielsen}}, \ and\
  \bibinfo {author} {\bibfnamefont {U.~L.}\ \bibnamefont {Andersen}},\
  }\href@noop {} {\bibfield  {journal} {\bibinfo  {journal} {Science}\ }\textbf
  {\bibinfo {volume} {366}},\ \bibinfo {pages} {369} (\bibinfo {year}
  {2019})}\BibitemShut {NoStop}%
\bibitem [{\citenamefont {Zhong}\ \emph {et~al.}(2020)\citenamefont {Zhong},
  \citenamefont {Wang}, \citenamefont {Deng}, \citenamefont {Chen},
  \citenamefont {Peng}, \citenamefont {Luo}, \citenamefont {Qin}, \citenamefont
  {Wu}, \citenamefont {Ding}, \citenamefont {Hu} \emph
  {et~al.}}]{zhong2020quantum}%
  \BibitemOpen
  \bibfield  {author} {\bibinfo {author} {\bibfnamefont {H.-S.}\ \bibnamefont
  {Zhong}}, \bibinfo {author} {\bibfnamefont {H.}~\bibnamefont {Wang}},
  \bibinfo {author} {\bibfnamefont {Y.-H.}\ \bibnamefont {Deng}}, \bibinfo
  {author} {\bibfnamefont {M.-C.}\ \bibnamefont {Chen}}, \bibinfo {author}
  {\bibfnamefont {L.-C.}\ \bibnamefont {Peng}}, \bibinfo {author}
  {\bibfnamefont {Y.-H.}\ \bibnamefont {Luo}}, \bibinfo {author} {\bibfnamefont
  {J.}~\bibnamefont {Qin}}, \bibinfo {author} {\bibfnamefont {D.}~\bibnamefont
  {Wu}}, \bibinfo {author} {\bibfnamefont {X.}~\bibnamefont {Ding}}, \bibinfo
  {author} {\bibfnamefont {Y.}~\bibnamefont {Hu}},  \emph {et~al.},\
  }\href@noop {} {\bibfield  {journal} {\bibinfo  {journal} {Science}\ }\textbf
  {\bibinfo {volume} {370}},\ \bibinfo {pages} {1460} (\bibinfo {year}
  {2020})}\BibitemShut {NoStop}%
\bibitem [{\citenamefont {Knill}\ \emph {et~al.}(2001)\citenamefont {Knill},
  \citenamefont {Laflamme},\ and\ \citenamefont {Milburn}}]{Knill:2001aa}%
  \BibitemOpen
  \bibfield  {author} {\bibinfo {author} {\bibfnamefont {E.}~\bibnamefont
  {Knill}}, \bibinfo {author} {\bibfnamefont {R.}~\bibnamefont {Laflamme}}, \
  and\ \bibinfo {author} {\bibfnamefont {G.~J.}\ \bibnamefont {Milburn}},\
  }\href {http://dx.doi.org/10.1038/35051009} {\bibfield  {journal} {\bibinfo
  {journal} {Nature}\ }\textbf {\bibinfo {volume} {409}},\ \bibinfo {pages}
  {46} (\bibinfo {year} {2001})}\BibitemShut {NoStop}%
\bibitem [{\citenamefont {Lloyd}\ and\ \citenamefont
  {Braunstein}(1999)}]{lloyd1999quantum}%
  \BibitemOpen
  \bibfield  {author} {\bibinfo {author} {\bibfnamefont {S.}~\bibnamefont
  {Lloyd}}\ and\ \bibinfo {author} {\bibfnamefont {S.~L.}\ \bibnamefont
  {Braunstein}},\ }\href@noop {} {\bibfield  {journal} {\bibinfo  {journal}
  {Physical Review Letters}\ }\textbf {\bibinfo {volume} {82}},\ \bibinfo
  {pages} {1784} (\bibinfo {year} {1999})}\BibitemShut {NoStop}%
\bibitem [{\citenamefont {Gu}\ \emph {et~al.}(2009)\citenamefont {Gu},
  \citenamefont {Weedbrook}, \citenamefont {Menicucci}, \citenamefont {Ralph},\
  and\ \citenamefont {van Loock}}]{PhysRevA.79.062318}%
  \BibitemOpen
  \bibfield  {author} {\bibinfo {author} {\bibfnamefont {M.}~\bibnamefont
  {Gu}}, \bibinfo {author} {\bibfnamefont {C.}~\bibnamefont {Weedbrook}},
  \bibinfo {author} {\bibfnamefont {N.~C.}\ \bibnamefont {Menicucci}}, \bibinfo
  {author} {\bibfnamefont {T.~C.}\ \bibnamefont {Ralph}}, \ and\ \bibinfo
  {author} {\bibfnamefont {P.}~\bibnamefont {van Loock}},\ }\href {\doibase
  10.1103/PhysRevA.79.062318} {\bibfield  {journal} {\bibinfo  {journal} {Phys.
  Rev. A}\ }\textbf {\bibinfo {volume} {79}},\ \bibinfo {pages} {062318}
  (\bibinfo {year} {2009})}\BibitemShut {NoStop}%
\bibitem [{\citenamefont {Yu}\ \emph {et~al.}(2020)\citenamefont {Yu},
  \citenamefont {McCuller}, \citenamefont {Tse}, \citenamefont {Kijbunchoo},
  \citenamefont {Barsotti},\ and\ \citenamefont {Mavalvala}}]{yu2020quantum}%
  \BibitemOpen
  \bibfield  {author} {\bibinfo {author} {\bibfnamefont {H.}~\bibnamefont
  {Yu}}, \bibinfo {author} {\bibfnamefont {L.}~\bibnamefont {McCuller}},
  \bibinfo {author} {\bibfnamefont {M.}~\bibnamefont {Tse}}, \bibinfo {author}
  {\bibfnamefont {N.}~\bibnamefont {Kijbunchoo}}, \bibinfo {author}
  {\bibfnamefont {L.}~\bibnamefont {Barsotti}}, \ and\ \bibinfo {author}
  {\bibfnamefont {N.}~\bibnamefont {Mavalvala}},\ }\href@noop {} {\bibfield
  {journal} {\bibinfo  {journal} {Nature}\ }\textbf {\bibinfo {volume} {583}},\
  \bibinfo {pages} {43} (\bibinfo {year} {2020})}\BibitemShut {NoStop}%
\bibitem [{\citenamefont {Casacio}\ \emph {et~al.}(2021)\citenamefont
  {Casacio}, \citenamefont {Madsen}, \citenamefont {Terrasson}, \citenamefont
  {Waleed}, \citenamefont {Barnscheidt}, \citenamefont {Hage}, \citenamefont
  {Taylor},\ and\ \citenamefont {Bowen}}]{casacio2021quantum}%
  \BibitemOpen
  \bibfield  {author} {\bibinfo {author} {\bibfnamefont {C.~A.}\ \bibnamefont
  {Casacio}}, \bibinfo {author} {\bibfnamefont {L.~S.}\ \bibnamefont {Madsen}},
  \bibinfo {author} {\bibfnamefont {A.}~\bibnamefont {Terrasson}}, \bibinfo
  {author} {\bibfnamefont {M.}~\bibnamefont {Waleed}}, \bibinfo {author}
  {\bibfnamefont {K.}~\bibnamefont {Barnscheidt}}, \bibinfo {author}
  {\bibfnamefont {B.}~\bibnamefont {Hage}}, \bibinfo {author} {\bibfnamefont
  {M.~A.}\ \bibnamefont {Taylor}}, \ and\ \bibinfo {author} {\bibfnamefont
  {W.~P.}\ \bibnamefont {Bowen}},\ }\href@noop {} {\bibfield  {journal}
  {\bibinfo  {journal} {Nature}\ }\textbf {\bibinfo {volume} {594}},\ \bibinfo
  {pages} {201} (\bibinfo {year} {2021})}\BibitemShut {NoStop}%
\bibitem [{\citenamefont {Furusawa}\ \emph {et~al.}(1998)\citenamefont
  {Furusawa}, \citenamefont {S{\o}rensen}, \citenamefont {Braunstein},
  \citenamefont {Fuchs}, \citenamefont {Kimble},\ and\ \citenamefont
  {Polzik}}]{furusawa1998unconditional}%
  \BibitemOpen
  \bibfield  {author} {\bibinfo {author} {\bibfnamefont {A.}~\bibnamefont
  {Furusawa}}, \bibinfo {author} {\bibfnamefont {J.~L.}\ \bibnamefont
  {S{\o}rensen}}, \bibinfo {author} {\bibfnamefont {S.~L.}\ \bibnamefont
  {Braunstein}}, \bibinfo {author} {\bibfnamefont {C.~A.}\ \bibnamefont
  {Fuchs}}, \bibinfo {author} {\bibfnamefont {H.~J.}\ \bibnamefont {Kimble}}, \
  and\ \bibinfo {author} {\bibfnamefont {E.~S.}\ \bibnamefont {Polzik}},\
  }\href@noop {} {\bibfield  {journal} {\bibinfo  {journal} {science}\ }\textbf
  {\bibinfo {volume} {282}},\ \bibinfo {pages} {706} (\bibinfo {year}
  {1998})}\BibitemShut {NoStop}%
\bibitem [{\citenamefont {Raussendorf}\ and\ \citenamefont
  {Briegel}(2001)}]{Raussendorf}%
  \BibitemOpen
  \bibfield  {author} {\bibinfo {author} {\bibfnamefont {R.}~\bibnamefont
  {Raussendorf}}\ and\ \bibinfo {author} {\bibfnamefont {H.~J.}\ \bibnamefont
  {Briegel}},\ }\href {\doibase 10.1103/PhysRevLett.86.5188} {\bibfield
  {journal} {\bibinfo  {journal} {Phys. Rev. Lett.}\ }\textbf {\bibinfo
  {volume} {86}},\ \bibinfo {pages} {5188} (\bibinfo {year}
  {2001})}\BibitemShut {NoStop}%
\bibitem [{\citenamefont {Kanter}\ \emph {et~al.}(2002)\citenamefont {Kanter},
  \citenamefont {Kumar}, \citenamefont {Roussev}, \citenamefont {Kurz},
  \citenamefont {Parameswaran},\ and\ \citenamefont
  {Fejer}}]{kanter2002squeezing}%
  \BibitemOpen
  \bibfield  {author} {\bibinfo {author} {\bibfnamefont {G.~S.}\ \bibnamefont
  {Kanter}}, \bibinfo {author} {\bibfnamefont {P.}~\bibnamefont {Kumar}},
  \bibinfo {author} {\bibfnamefont {R.~V.}\ \bibnamefont {Roussev}}, \bibinfo
  {author} {\bibfnamefont {J.}~\bibnamefont {Kurz}}, \bibinfo {author}
  {\bibfnamefont {K.~R.}\ \bibnamefont {Parameswaran}}, \ and\ \bibinfo
  {author} {\bibfnamefont {M.~M.}\ \bibnamefont {Fejer}},\ }\href@noop {}
  {\bibfield  {journal} {\bibinfo  {journal} {Optics express}\ }\textbf
  {\bibinfo {volume} {10}},\ \bibinfo {pages} {177} (\bibinfo {year}
  {2002})}\BibitemShut {NoStop}%
\bibitem [{\citenamefont {Larsen}\ \emph {et~al.}(2020)\citenamefont {Larsen},
  \citenamefont {Guo}, \citenamefont {Breum}, \citenamefont
  {Neergaard-Nielsen},\ and\ \citenamefont
  {Andersen}}]{larsen2020deterministic}%
  \BibitemOpen
  \bibfield  {author} {\bibinfo {author} {\bibfnamefont {M.~V.}\ \bibnamefont
  {Larsen}}, \bibinfo {author} {\bibfnamefont {X.}~\bibnamefont {Guo}},
  \bibinfo {author} {\bibfnamefont {C.~R.}\ \bibnamefont {Breum}}, \bibinfo
  {author} {\bibfnamefont {J.~S.}\ \bibnamefont {Neergaard-Nielsen}}, \ and\
  \bibinfo {author} {\bibfnamefont {U.~L.}\ \bibnamefont {Andersen}},\
  }\href@noop {} {\bibfield  {journal} {\bibinfo  {journal} {arXiv preprint
  arXiv:2010.14422}\ } (\bibinfo {year} {2020})}\BibitemShut {NoStop}%
\bibitem [{\citenamefont {Joshi}\ \emph {et~al.}(2020)\citenamefont {Joshi},
  \citenamefont {Farsi}, \citenamefont {Dutt}, \citenamefont {Kim},
  \citenamefont {Ji}, \citenamefont {Zhao}, \citenamefont {Bishop},
  \citenamefont {Lipson},\ and\ \citenamefont {Gaeta}}]{joshi2020frequency}%
  \BibitemOpen
  \bibfield  {author} {\bibinfo {author} {\bibfnamefont {C.}~\bibnamefont
  {Joshi}}, \bibinfo {author} {\bibfnamefont {A.}~\bibnamefont {Farsi}},
  \bibinfo {author} {\bibfnamefont {A.}~\bibnamefont {Dutt}}, \bibinfo {author}
  {\bibfnamefont {B.~Y.}\ \bibnamefont {Kim}}, \bibinfo {author} {\bibfnamefont
  {X.}~\bibnamefont {Ji}}, \bibinfo {author} {\bibfnamefont {Y.}~\bibnamefont
  {Zhao}}, \bibinfo {author} {\bibfnamefont {A.~M.}\ \bibnamefont {Bishop}},
  \bibinfo {author} {\bibfnamefont {M.}~\bibnamefont {Lipson}}, \ and\ \bibinfo
  {author} {\bibfnamefont {A.~L.}\ \bibnamefont {Gaeta}},\ }\href@noop {}
  {\bibfield  {journal} {\bibinfo  {journal} {Physical review letters}\
  }\textbf {\bibinfo {volume} {124}},\ \bibinfo {pages} {143601} (\bibinfo
  {year} {2020})}\BibitemShut {NoStop}%
\bibitem [{\citenamefont {Vaidya}\ \emph {et~al.}(2020)\citenamefont {Vaidya},
  \citenamefont {Morrison}, \citenamefont {Helt}, \citenamefont {Shahrokshahi},
  \citenamefont {Mahler}, \citenamefont {Collins}, \citenamefont {Tan},
  \citenamefont {Lavoie}, \citenamefont {Repingon}, \citenamefont {Menotti}
  \emph {et~al.}}]{vaidya2020broadband}%
  \BibitemOpen
  \bibfield  {author} {\bibinfo {author} {\bibfnamefont {V.}~\bibnamefont
  {Vaidya}}, \bibinfo {author} {\bibfnamefont {B.}~\bibnamefont {Morrison}},
  \bibinfo {author} {\bibfnamefont {L.}~\bibnamefont {Helt}}, \bibinfo {author}
  {\bibfnamefont {R.}~\bibnamefont {Shahrokshahi}}, \bibinfo {author}
  {\bibfnamefont {D.}~\bibnamefont {Mahler}}, \bibinfo {author} {\bibfnamefont
  {M.}~\bibnamefont {Collins}}, \bibinfo {author} {\bibfnamefont
  {K.}~\bibnamefont {Tan}}, \bibinfo {author} {\bibfnamefont {J.}~\bibnamefont
  {Lavoie}}, \bibinfo {author} {\bibfnamefont {A.}~\bibnamefont {Repingon}},
  \bibinfo {author} {\bibfnamefont {M.}~\bibnamefont {Menotti}},  \emph
  {et~al.},\ }\href@noop {} {\bibfield  {journal} {\bibinfo  {journal} {Science
  advances}\ }\textbf {\bibinfo {volume} {6}},\ \bibinfo {pages} {eaba9186}
  (\bibinfo {year} {2020})}\BibitemShut {NoStop}%
\bibitem [{\citenamefont {Yang}\ \emph {et~al.}(2021)\citenamefont {Yang},
  \citenamefont {Jahanbozorgi}, \citenamefont {Jeong}, \citenamefont {Sun},
  \citenamefont {Pfister}, \citenamefont {Lee},\ and\ \citenamefont
  {Yi}}]{yang2021squeezed}%
  \BibitemOpen
  \bibfield  {author} {\bibinfo {author} {\bibfnamefont {Z.}~\bibnamefont
  {Yang}}, \bibinfo {author} {\bibfnamefont {M.}~\bibnamefont {Jahanbozorgi}},
  \bibinfo {author} {\bibfnamefont {D.}~\bibnamefont {Jeong}}, \bibinfo
  {author} {\bibfnamefont {S.}~\bibnamefont {Sun}}, \bibinfo {author}
  {\bibfnamefont {O.}~\bibnamefont {Pfister}}, \bibinfo {author} {\bibfnamefont
  {H.}~\bibnamefont {Lee}}, \ and\ \bibinfo {author} {\bibfnamefont
  {X.}~\bibnamefont {Yi}},\ }\href@noop {} {\bibfield  {journal} {\bibinfo
  {journal} {Nature Communications}\ }\textbf {\bibinfo {volume} {12}},\
  \bibinfo {pages} {1} (\bibinfo {year} {2021})}\BibitemShut {NoStop}%
\bibitem [{\citenamefont {Vigliar}\ \emph {et~al.}(2021)\citenamefont
  {Vigliar}, \citenamefont {Paesani}, \citenamefont {Ding}, \citenamefont
  {Adcock}, \citenamefont {Wang}, \citenamefont {Morley-Short}, \citenamefont
  {Bacco}, \citenamefont {Oxenl{\o}we}, \citenamefont {Thompson}, \citenamefont
  {Rarity} \emph {et~al.}}]{vigliar2021error}%
  \BibitemOpen
  \bibfield  {author} {\bibinfo {author} {\bibfnamefont {C.}~\bibnamefont
  {Vigliar}}, \bibinfo {author} {\bibfnamefont {S.}~\bibnamefont {Paesani}},
  \bibinfo {author} {\bibfnamefont {Y.}~\bibnamefont {Ding}}, \bibinfo {author}
  {\bibfnamefont {J.~C.}\ \bibnamefont {Adcock}}, \bibinfo {author}
  {\bibfnamefont {J.}~\bibnamefont {Wang}}, \bibinfo {author} {\bibfnamefont
  {S.}~\bibnamefont {Morley-Short}}, \bibinfo {author} {\bibfnamefont
  {D.}~\bibnamefont {Bacco}}, \bibinfo {author} {\bibfnamefont {L.~K.}\
  \bibnamefont {Oxenl{\o}we}}, \bibinfo {author} {\bibfnamefont {M.~G.}\
  \bibnamefont {Thompson}}, \bibinfo {author} {\bibfnamefont {J.~G.}\
  \bibnamefont {Rarity}},  \emph {et~al.},\ }\href@noop {} {\bibfield
  {journal} {\bibinfo  {journal} {Nature Physics}\ }\textbf {\bibinfo {volume}
  {17}},\ \bibinfo {pages} {1137} (\bibinfo {year} {2021})}\BibitemShut
  {NoStop}%
\bibitem [{\citenamefont {Arrazola}\ \emph {et~al.}(2021)\citenamefont
  {Arrazola}, \citenamefont {Bergholm}, \citenamefont {Br{\'a}dler},
  \citenamefont {Bromley}, \citenamefont {Collins}, \citenamefont {Dhand},
  \citenamefont {Fumagalli}, \citenamefont {Gerrits}, \citenamefont {Goussev},
  \citenamefont {Helt} \emph {et~al.}}]{arrazola2021quantum}%
  \BibitemOpen
  \bibfield  {author} {\bibinfo {author} {\bibfnamefont {J.}~\bibnamefont
  {Arrazola}}, \bibinfo {author} {\bibfnamefont {V.}~\bibnamefont {Bergholm}},
  \bibinfo {author} {\bibfnamefont {K.}~\bibnamefont {Br{\'a}dler}}, \bibinfo
  {author} {\bibfnamefont {T.}~\bibnamefont {Bromley}}, \bibinfo {author}
  {\bibfnamefont {M.}~\bibnamefont {Collins}}, \bibinfo {author} {\bibfnamefont
  {I.}~\bibnamefont {Dhand}}, \bibinfo {author} {\bibfnamefont
  {A.}~\bibnamefont {Fumagalli}}, \bibinfo {author} {\bibfnamefont
  {T.}~\bibnamefont {Gerrits}}, \bibinfo {author} {\bibfnamefont
  {A.}~\bibnamefont {Goussev}}, \bibinfo {author} {\bibfnamefont
  {L.}~\bibnamefont {Helt}},  \emph {et~al.},\ }\href@noop {} {\bibfield
  {journal} {\bibinfo  {journal} {Nature}\ }\textbf {\bibinfo {volume} {591}},\
  \bibinfo {pages} {54} (\bibinfo {year} {2021})}\BibitemShut {NoStop}%
\bibitem [{\citenamefont {Dutt}\ \emph {et~al.}(2015)\citenamefont {Dutt},
  \citenamefont {Luke}, \citenamefont {Manipatruni}, \citenamefont {Gaeta},
  \citenamefont {Nussenzveig},\ and\ \citenamefont {Lipson}}]{dutt2015chip}%
  \BibitemOpen
  \bibfield  {author} {\bibinfo {author} {\bibfnamefont {A.}~\bibnamefont
  {Dutt}}, \bibinfo {author} {\bibfnamefont {K.}~\bibnamefont {Luke}}, \bibinfo
  {author} {\bibfnamefont {S.}~\bibnamefont {Manipatruni}}, \bibinfo {author}
  {\bibfnamefont {A.~L.}\ \bibnamefont {Gaeta}}, \bibinfo {author}
  {\bibfnamefont {P.}~\bibnamefont {Nussenzveig}}, \ and\ \bibinfo {author}
  {\bibfnamefont {M.}~\bibnamefont {Lipson}},\ }\href@noop {} {\bibfield
  {journal} {\bibinfo  {journal} {Physical Review Applied}\ }\textbf {\bibinfo
  {volume} {3}},\ \bibinfo {pages} {044005} (\bibinfo {year}
  {2015})}\BibitemShut {NoStop}%
\bibitem [{\citenamefont {Cernansky}\ and\ \citenamefont
  {Politi}(2020)}]{cernansky2020nanophotonic}%
  \BibitemOpen
  \bibfield  {author} {\bibinfo {author} {\bibfnamefont {R.}~\bibnamefont
  {Cernansky}}\ and\ \bibinfo {author} {\bibfnamefont {A.}~\bibnamefont
  {Politi}},\ }\href@noop {} {\bibfield  {journal} {\bibinfo  {journal} {APL
  Photonics}\ }\textbf {\bibinfo {volume} {5}},\ \bibinfo {pages} {101303}
  (\bibinfo {year} {2020})}\BibitemShut {NoStop}%
\bibitem [{\citenamefont {Tasker}\ \emph {et~al.}(2021)\citenamefont {Tasker},
  \citenamefont {Frazer}, \citenamefont {Ferranti}, \citenamefont {Allen},
  \citenamefont {Brunel}, \citenamefont {Tanzilli}, \citenamefont {D’Auria},\
  and\ \citenamefont {Matthews}}]{tasker2021silicon}%
  \BibitemOpen
  \bibfield  {author} {\bibinfo {author} {\bibfnamefont {J.~F.}\ \bibnamefont
  {Tasker}}, \bibinfo {author} {\bibfnamefont {J.}~\bibnamefont {Frazer}},
  \bibinfo {author} {\bibfnamefont {G.}~\bibnamefont {Ferranti}}, \bibinfo
  {author} {\bibfnamefont {E.~J.}\ \bibnamefont {Allen}}, \bibinfo {author}
  {\bibfnamefont {L.~F.}\ \bibnamefont {Brunel}}, \bibinfo {author}
  {\bibfnamefont {S.}~\bibnamefont {Tanzilli}}, \bibinfo {author}
  {\bibfnamefont {V.}~\bibnamefont {D’Auria}}, \ and\ \bibinfo {author}
  {\bibfnamefont {J.~C.}\ \bibnamefont {Matthews}},\ }\href@noop {} {\bibfield
  {journal} {\bibinfo  {journal} {Nature Photonics}\ }\textbf {\bibinfo
  {volume} {15}},\ \bibinfo {pages} {11} (\bibinfo {year} {2021})}\BibitemShut
  {NoStop}%
\bibitem [{\citenamefont {Zhao}\ \emph {et~al.}(2020)\citenamefont {Zhao},
  \citenamefont {Ma}, \citenamefont {R\"using},\ and\ \citenamefont
  {Mookherjea}}]{Shayan_PRL}%
  \BibitemOpen
  \bibfield  {author} {\bibinfo {author} {\bibfnamefont {J.}~\bibnamefont
  {Zhao}}, \bibinfo {author} {\bibfnamefont {C.}~\bibnamefont {Ma}}, \bibinfo
  {author} {\bibfnamefont {M.}~\bibnamefont {R\"using}}, \ and\ \bibinfo
  {author} {\bibfnamefont {S.}~\bibnamefont {Mookherjea}},\ }\href {\doibase
  10.1103/PhysRevLett.124.163603} {\bibfield  {journal} {\bibinfo  {journal}
  {Phys. Rev. Lett.}\ }\textbf {\bibinfo {volume} {124}},\ \bibinfo {pages}
  {163603} (\bibinfo {year} {2020})}\BibitemShut {NoStop}%
\bibitem [{\citenamefont {Javid}\ \emph {et~al.}(2021)\citenamefont {Javid},
  \citenamefont {Ling}, \citenamefont {Staffa}, \citenamefont {Li},
  \citenamefont {He},\ and\ \citenamefont {Lin}}]{PhysRevLett.127.183601}%
  \BibitemOpen
  \bibfield  {author} {\bibinfo {author} {\bibfnamefont {U.~A.}\ \bibnamefont
  {Javid}}, \bibinfo {author} {\bibfnamefont {J.}~\bibnamefont {Ling}},
  \bibinfo {author} {\bibfnamefont {J.}~\bibnamefont {Staffa}}, \bibinfo
  {author} {\bibfnamefont {M.}~\bibnamefont {Li}}, \bibinfo {author}
  {\bibfnamefont {Y.}~\bibnamefont {He}}, \ and\ \bibinfo {author}
  {\bibfnamefont {Q.}~\bibnamefont {Lin}},\ }\href {\doibase
  10.1103/PhysRevLett.127.183601} {\bibfield  {journal} {\bibinfo  {journal}
  {Phys. Rev. Lett.}\ }\textbf {\bibinfo {volume} {127}},\ \bibinfo {pages}
  {183601} (\bibinfo {year} {2021})}\BibitemShut {NoStop}%
\bibitem [{\citenamefont {Chen}\ \emph {et~al.}(2021)\citenamefont {Chen},
  \citenamefont {Briggs}, \citenamefont {Hou},\ and\ \citenamefont
  {Fan}}]{chen2021ultra}%
  \BibitemOpen
  \bibfield  {author} {\bibinfo {author} {\bibfnamefont {P.-K.}\ \bibnamefont
  {Chen}}, \bibinfo {author} {\bibfnamefont {I.}~\bibnamefont {Briggs}},
  \bibinfo {author} {\bibfnamefont {S.}~\bibnamefont {Hou}}, \ and\ \bibinfo
  {author} {\bibfnamefont {L.}~\bibnamefont {Fan}},\ }\href@noop {} {\bibfield
  {journal} {\bibinfo  {journal} {arXiv preprint arXiv:2107.02250}\ } (\bibinfo
  {year} {2021})}\BibitemShut {NoStop}%
\bibitem [{\citenamefont {Caves}(1982)}]{caves1982quantum}%
  \BibitemOpen
  \bibfield  {author} {\bibinfo {author} {\bibfnamefont {C.~M.}\ \bibnamefont
  {Caves}},\ }\href@noop {} {\bibfield  {journal} {\bibinfo  {journal}
  {Physical Review D}\ }\textbf {\bibinfo {volume} {26}},\ \bibinfo {pages}
  {1817} (\bibinfo {year} {1982})}\BibitemShut {NoStop}%
\bibitem [{\citenamefont {Shaked}\ \emph {et~al.}(2018)\citenamefont {Shaked},
  \citenamefont {Michael}, \citenamefont {Vered}, \citenamefont {Bello},
  \citenamefont {Rosenbluh},\ and\ \citenamefont
  {Pe’er}}]{shaked2018lifting}%
  \BibitemOpen
  \bibfield  {author} {\bibinfo {author} {\bibfnamefont {Y.}~\bibnamefont
  {Shaked}}, \bibinfo {author} {\bibfnamefont {Y.}~\bibnamefont {Michael}},
  \bibinfo {author} {\bibfnamefont {R.~Z.}\ \bibnamefont {Vered}}, \bibinfo
  {author} {\bibfnamefont {L.}~\bibnamefont {Bello}}, \bibinfo {author}
  {\bibfnamefont {M.}~\bibnamefont {Rosenbluh}}, \ and\ \bibinfo {author}
  {\bibfnamefont {A.}~\bibnamefont {Pe’er}},\ }\href@noop {} {\bibfield
  {journal} {\bibinfo  {journal} {Nature communications}\ }\textbf {\bibinfo
  {volume} {9}},\ \bibinfo {pages} {1} (\bibinfo {year} {2018})}\BibitemShut
  {NoStop}%
\bibitem [{\citenamefont {Takanashi}\ \emph {et~al.}(2020)\citenamefont
  {Takanashi}, \citenamefont {Inoue}, \citenamefont {Kashiwazaki},
  \citenamefont {Kazama}, \citenamefont {Enbutsu}, \citenamefont {Kasahara},
  \citenamefont {Umeki},\ and\ \citenamefont {Furusawa}}]{takanashi2020all}%
  \BibitemOpen
  \bibfield  {author} {\bibinfo {author} {\bibfnamefont {N.}~\bibnamefont
  {Takanashi}}, \bibinfo {author} {\bibfnamefont {A.}~\bibnamefont {Inoue}},
  \bibinfo {author} {\bibfnamefont {T.}~\bibnamefont {Kashiwazaki}}, \bibinfo
  {author} {\bibfnamefont {T.}~\bibnamefont {Kazama}}, \bibinfo {author}
  {\bibfnamefont {K.}~\bibnamefont {Enbutsu}}, \bibinfo {author} {\bibfnamefont
  {R.}~\bibnamefont {Kasahara}}, \bibinfo {author} {\bibfnamefont
  {T.}~\bibnamefont {Umeki}}, \ and\ \bibinfo {author} {\bibfnamefont
  {A.}~\bibnamefont {Furusawa}},\ }\href@noop {} {\bibfield  {journal}
  {\bibinfo  {journal} {Optics Express}\ }\textbf {\bibinfo {volume} {28}},\
  \bibinfo {pages} {34916} (\bibinfo {year} {2020})}\BibitemShut {NoStop}%
\bibitem [{\citenamefont {Frascella}\ \emph {et~al.}(2021)\citenamefont
  {Frascella}, \citenamefont {Agne}, \citenamefont {Khalili},\ and\
  \citenamefont {Chekhova}}]{frascella2021overcoming}%
  \BibitemOpen
  \bibfield  {author} {\bibinfo {author} {\bibfnamefont {G.}~\bibnamefont
  {Frascella}}, \bibinfo {author} {\bibfnamefont {S.}~\bibnamefont {Agne}},
  \bibinfo {author} {\bibfnamefont {F.~Y.}\ \bibnamefont {Khalili}}, \ and\
  \bibinfo {author} {\bibfnamefont {M.~V.}\ \bibnamefont {Chekhova}},\
  }\href@noop {} {\bibfield  {journal} {\bibinfo  {journal} {npj Quantum
  Information}\ }\textbf {\bibinfo {volume} {7}},\ \bibinfo {pages} {1}
  (\bibinfo {year} {2021})}\BibitemShut {NoStop}%
\bibitem [{\citenamefont {Li}\ \emph {et~al.}(2020)\citenamefont {Li},
  \citenamefont {Liu}, \citenamefont {Huo}, \citenamefont {Cui}, \citenamefont
  {Feng}, \citenamefont {Li},\ and\ \citenamefont {Ou}}]{PhysRevA.101.053801}%
  \BibitemOpen
  \bibfield  {author} {\bibinfo {author} {\bibfnamefont {J.}~\bibnamefont
  {Li}}, \bibinfo {author} {\bibfnamefont {Y.}~\bibnamefont {Liu}}, \bibinfo
  {author} {\bibfnamefont {N.}~\bibnamefont {Huo}}, \bibinfo {author}
  {\bibfnamefont {L.}~\bibnamefont {Cui}}, \bibinfo {author} {\bibfnamefont
  {S.}~\bibnamefont {Feng}}, \bibinfo {author} {\bibfnamefont {X.}~\bibnamefont
  {Li}}, \ and\ \bibinfo {author} {\bibfnamefont {Z.~Y.}\ \bibnamefont {Ou}},\
  }\href {\doibase 10.1103/PhysRevA.101.053801} {\bibfield  {journal} {\bibinfo
   {journal} {Phys. Rev. A}\ }\textbf {\bibinfo {volume} {101}},\ \bibinfo
  {pages} {053801} (\bibinfo {year} {2020})}\BibitemShut {NoStop}%
\bibitem [{\citenamefont {Ye}\ \emph {et~al.}(2021)\citenamefont {Ye},
  \citenamefont {Zhao}, \citenamefont {Twayana}, \citenamefont {Karlsson},
  \citenamefont {Torres-Company},\ and\ \citenamefont
  {Andrekson}}]{ye2021overcoming}%
  \BibitemOpen
  \bibfield  {author} {\bibinfo {author} {\bibfnamefont {Z.}~\bibnamefont
  {Ye}}, \bibinfo {author} {\bibfnamefont {P.}~\bibnamefont {Zhao}}, \bibinfo
  {author} {\bibfnamefont {K.}~\bibnamefont {Twayana}}, \bibinfo {author}
  {\bibfnamefont {M.}~\bibnamefont {Karlsson}}, \bibinfo {author}
  {\bibfnamefont {V.}~\bibnamefont {Torres-Company}}, \ and\ \bibinfo {author}
  {\bibfnamefont {P.~A.}\ \bibnamefont {Andrekson}},\ }\href@noop {} {\bibfield
   {journal} {\bibinfo  {journal} {Science advances}\ }\textbf {\bibinfo
  {volume} {7}},\ \bibinfo {pages} {eabi8150} (\bibinfo {year}
  {2021})}\BibitemShut {NoStop}%
\bibitem [{\citenamefont {Zhu}\ \emph {et~al.}(2021)\citenamefont {Zhu},
  \citenamefont {Shao}, \citenamefont {Yu}, \citenamefont {Cheng},
  \citenamefont {Desiatov}, \citenamefont {Xin}, \citenamefont {Hu},
  \citenamefont {Holzgrafe}, \citenamefont {Ghosh}, \citenamefont
  {Shams-Ansari} \emph {et~al.}}]{zhu2021integrated}%
  \BibitemOpen
  \bibfield  {author} {\bibinfo {author} {\bibfnamefont {D.}~\bibnamefont
  {Zhu}}, \bibinfo {author} {\bibfnamefont {L.}~\bibnamefont {Shao}}, \bibinfo
  {author} {\bibfnamefont {M.}~\bibnamefont {Yu}}, \bibinfo {author}
  {\bibfnamefont {R.}~\bibnamefont {Cheng}}, \bibinfo {author} {\bibfnamefont
  {B.}~\bibnamefont {Desiatov}}, \bibinfo {author} {\bibfnamefont
  {C.}~\bibnamefont {Xin}}, \bibinfo {author} {\bibfnamefont {Y.}~\bibnamefont
  {Hu}}, \bibinfo {author} {\bibfnamefont {J.}~\bibnamefont {Holzgrafe}},
  \bibinfo {author} {\bibfnamefont {S.}~\bibnamefont {Ghosh}}, \bibinfo
  {author} {\bibfnamefont {A.}~\bibnamefont {Shams-Ansari}},  \emph {et~al.},\
  }\href@noop {} {\bibfield  {journal} {\bibinfo  {journal} {Advances in Optics
  and Photonics}\ }\textbf {\bibinfo {volume} {13}},\ \bibinfo {pages} {242}
  (\bibinfo {year} {2021})}\BibitemShut {NoStop}%
\bibitem [{\citenamefont {Jankowski}\ \emph {et~al.}(2021)\citenamefont
  {Jankowski}, \citenamefont {Mishra},\ and\ \citenamefont
  {Fejer}}]{jankowski2021dispersion}%
  \BibitemOpen
  \bibfield  {author} {\bibinfo {author} {\bibfnamefont {M.}~\bibnamefont
  {Jankowski}}, \bibinfo {author} {\bibfnamefont {J.}~\bibnamefont {Mishra}}, \
  and\ \bibinfo {author} {\bibfnamefont {M.~M.}\ \bibnamefont {Fejer}},\
  }\href@noop {} {\bibfield  {journal} {\bibinfo  {journal} {Journal of
  Physics: Photonics}\ } (\bibinfo {year} {2021})}\BibitemShut {NoStop}%
\bibitem [{\citenamefont {Ledezma}\ \emph {et~al.}(2021)\citenamefont
  {Ledezma}, \citenamefont {Sekine}, \citenamefont {Guo}, \citenamefont
  {Nehra}, \citenamefont {Jahani},\ and\ \citenamefont
  {Marandi}}]{ledezma2021intense}%
  \BibitemOpen
  \bibfield  {author} {\bibinfo {author} {\bibfnamefont {L.}~\bibnamefont
  {Ledezma}}, \bibinfo {author} {\bibfnamefont {R.}~\bibnamefont {Sekine}},
  \bibinfo {author} {\bibfnamefont {Q.}~\bibnamefont {Guo}}, \bibinfo {author}
  {\bibfnamefont {R.}~\bibnamefont {Nehra}}, \bibinfo {author} {\bibfnamefont
  {S.}~\bibnamefont {Jahani}}, \ and\ \bibinfo {author} {\bibfnamefont
  {A.}~\bibnamefont {Marandi}},\ }\href@noop {} {\bibfield  {journal} {\bibinfo
   {journal} {arXiv preprint arXiv:2104.08262}\ } (\bibinfo {year}
  {2021})}\BibitemShut {NoStop}%
\bibitem [{\citenamefont {Jankowski}\ \emph {et~al.}(2020)\citenamefont
  {Jankowski}, \citenamefont {Langrock}, \citenamefont {Desiatov},
  \citenamefont {Marandi}, \citenamefont {Wang}, \citenamefont {Zhang},
  \citenamefont {Phillips}, \citenamefont {Lon\v{c}ar},\ and\ \citenamefont
  {Fejer}}]{Jankowski:20}%
  \BibitemOpen
  \bibfield  {author} {\bibinfo {author} {\bibfnamefont {M.}~\bibnamefont
  {Jankowski}}, \bibinfo {author} {\bibfnamefont {C.}~\bibnamefont {Langrock}},
  \bibinfo {author} {\bibfnamefont {B.}~\bibnamefont {Desiatov}}, \bibinfo
  {author} {\bibfnamefont {A.}~\bibnamefont {Marandi}}, \bibinfo {author}
  {\bibfnamefont {C.}~\bibnamefont {Wang}}, \bibinfo {author} {\bibfnamefont
  {M.}~\bibnamefont {Zhang}}, \bibinfo {author} {\bibfnamefont {C.~R.}\
  \bibnamefont {Phillips}}, \bibinfo {author} {\bibfnamefont {M.}~\bibnamefont
  {Lon\v{c}ar}}, \ and\ \bibinfo {author} {\bibfnamefont {M.~M.}\ \bibnamefont
  {Fejer}},\ }\href {\doibase 10.1364/OPTICA.7.000040} {\bibfield  {journal}
  {\bibinfo  {journal} {Optica}\ }\textbf {\bibinfo {volume} {7}},\ \bibinfo
  {pages} {40} (\bibinfo {year} {2020})}\BibitemShut {NoStop}%
\bibitem [{\citenamefont {Guo}\ \emph {et~al.}(2021)\citenamefont {Guo},
  \citenamefont {Sekine}, \citenamefont {Ledezma}, \citenamefont {Nehra},
  \citenamefont {Dean}, \citenamefont {Roy}, \citenamefont {Gray},
  \citenamefont {Jahani},\ and\ \citenamefont {Marandi}}]{guo2021femtojoule}%
  \BibitemOpen
  \bibfield  {author} {\bibinfo {author} {\bibfnamefont {Q.}~\bibnamefont
  {Guo}}, \bibinfo {author} {\bibfnamefont {R.}~\bibnamefont {Sekine}},
  \bibinfo {author} {\bibfnamefont {L.}~\bibnamefont {Ledezma}}, \bibinfo
  {author} {\bibfnamefont {R.}~\bibnamefont {Nehra}}, \bibinfo {author}
  {\bibfnamefont {D.~J.}\ \bibnamefont {Dean}}, \bibinfo {author}
  {\bibfnamefont {A.}~\bibnamefont {Roy}}, \bibinfo {author} {\bibfnamefont
  {R.~M.}\ \bibnamefont {Gray}}, \bibinfo {author} {\bibfnamefont
  {S.}~\bibnamefont {Jahani}}, \ and\ \bibinfo {author} {\bibfnamefont
  {A.}~\bibnamefont {Marandi}},\ }\href@noop {} {\bibfield  {journal} {\bibinfo
   {journal} {arXiv preprint arXiv:2107.09906}\ } (\bibinfo {year}
  {2021})}\BibitemShut {NoStop}%
\bibitem [{\citenamefont {Hu}\ \emph {et~al.}(2021)\citenamefont {Hu},
  \citenamefont {Yu}, \citenamefont {Zhu}, \citenamefont {Sinclair},
  \citenamefont {Shams-Ansari}, \citenamefont {Shao}, \citenamefont
  {Holzgrafe}, \citenamefont {Puma}, \citenamefont {Zhang},\ and\ \citenamefont
  {Lon{\v{c}}ar}}]{hu2021chip}%
  \BibitemOpen
  \bibfield  {author} {\bibinfo {author} {\bibfnamefont {Y.}~\bibnamefont
  {Hu}}, \bibinfo {author} {\bibfnamefont {M.}~\bibnamefont {Yu}}, \bibinfo
  {author} {\bibfnamefont {D.}~\bibnamefont {Zhu}}, \bibinfo {author}
  {\bibfnamefont {N.}~\bibnamefont {Sinclair}}, \bibinfo {author}
  {\bibfnamefont {A.}~\bibnamefont {Shams-Ansari}}, \bibinfo {author}
  {\bibfnamefont {L.}~\bibnamefont {Shao}}, \bibinfo {author} {\bibfnamefont
  {J.}~\bibnamefont {Holzgrafe}}, \bibinfo {author} {\bibfnamefont
  {E.}~\bibnamefont {Puma}}, \bibinfo {author} {\bibfnamefont {M.}~\bibnamefont
  {Zhang}}, \ and\ \bibinfo {author} {\bibfnamefont {M.}~\bibnamefont
  {Lon{\v{c}}ar}},\ }\href@noop {} {\bibfield  {journal} {\bibinfo  {journal}
  {Nature}\ }\textbf {\bibinfo {volume} {599}},\ \bibinfo {pages} {587}
  (\bibinfo {year} {2021})}\BibitemShut {NoStop}%
\bibitem [{Sup()}]{SuppMat}%
  \BibitemOpen
  \href@noop {} {\bibinfo  {journal} {{See supplementary information}}\
  }\BibitemShut {NoStop}%
\bibitem [{\citenamefont {Werner}\ \emph {et~al.}(1995)\citenamefont {Werner},
  \citenamefont {Raymer}, \citenamefont {Beck},\ and\ \citenamefont
  {Drummond}}]{PhysRevA.52.4202}%
  \BibitemOpen
\bibfield  {journal} {  }\bibfield  {author} {\bibinfo {author} {\bibfnamefont
  {M.~J.}\ \bibnamefont {Werner}}, \bibinfo {author} {\bibfnamefont {M.~G.}\
  \bibnamefont {Raymer}}, \bibinfo {author} {\bibfnamefont {M.}~\bibnamefont
  {Beck}}, \ and\ \bibinfo {author} {\bibfnamefont {P.~D.}\ \bibnamefont
  {Drummond}},\ }\href {\doibase 10.1103/PhysRevA.52.4202} {\bibfield
  {journal} {\bibinfo  {journal} {Phys. Rev. A}\ }\textbf {\bibinfo {volume}
  {52}},\ \bibinfo {pages} {4202} (\bibinfo {year} {1995})}\BibitemShut
  {NoStop}%
\bibitem [{\citenamefont {Oelker}\ \emph {et~al.}(2016)\citenamefont {Oelker},
  \citenamefont {Mansell}, \citenamefont {Tse}, \citenamefont {Miller},
  \citenamefont {Matichard}, \citenamefont {Barsotti}, \citenamefont
  {Fritschel}, \citenamefont {McClelland}, \citenamefont {Evans},\ and\
  \citenamefont {Mavalvala}}]{oelker2016ultra}%
  \BibitemOpen
  \bibfield  {author} {\bibinfo {author} {\bibfnamefont {E.}~\bibnamefont
  {Oelker}}, \bibinfo {author} {\bibfnamefont {G.}~\bibnamefont {Mansell}},
  \bibinfo {author} {\bibfnamefont {M.}~\bibnamefont {Tse}}, \bibinfo {author}
  {\bibfnamefont {J.}~\bibnamefont {Miller}}, \bibinfo {author} {\bibfnamefont
  {F.}~\bibnamefont {Matichard}}, \bibinfo {author} {\bibfnamefont
  {L.}~\bibnamefont {Barsotti}}, \bibinfo {author} {\bibfnamefont
  {P.}~\bibnamefont {Fritschel}}, \bibinfo {author} {\bibfnamefont
  {D.}~\bibnamefont {McClelland}}, \bibinfo {author} {\bibfnamefont
  {M.}~\bibnamefont {Evans}}, \ and\ \bibinfo {author} {\bibfnamefont
  {N.}~\bibnamefont {Mavalvala}},\ }\href@noop {} {\bibfield  {journal}
  {\bibinfo  {journal} {Optica}\ }\textbf {\bibinfo {volume} {3}},\ \bibinfo
  {pages} {682} (\bibinfo {year} {2016})}\BibitemShut {NoStop}%
\bibitem [{\citenamefont {Bourassa}\ \emph {et~al.}(2021)\citenamefont
  {Bourassa}, \citenamefont {Alexander}, \citenamefont {Vasmer}, \citenamefont
  {Patil}, \citenamefont {Tzitrin}, \citenamefont {Matsuura}, \citenamefont
  {Su}, \citenamefont {Baragiola}, \citenamefont {Guha}, \citenamefont
  {Dauphinais} \emph {et~al.}}]{bourassa2021blueprint}%
  \BibitemOpen
  \bibfield  {author} {\bibinfo {author} {\bibfnamefont {J.~E.}\ \bibnamefont
  {Bourassa}}, \bibinfo {author} {\bibfnamefont {R.~N.}\ \bibnamefont
  {Alexander}}, \bibinfo {author} {\bibfnamefont {M.}~\bibnamefont {Vasmer}},
  \bibinfo {author} {\bibfnamefont {A.}~\bibnamefont {Patil}}, \bibinfo
  {author} {\bibfnamefont {I.}~\bibnamefont {Tzitrin}}, \bibinfo {author}
  {\bibfnamefont {T.}~\bibnamefont {Matsuura}}, \bibinfo {author}
  {\bibfnamefont {D.}~\bibnamefont {Su}}, \bibinfo {author} {\bibfnamefont
  {B.~Q.}\ \bibnamefont {Baragiola}}, \bibinfo {author} {\bibfnamefont
  {S.}~\bibnamefont {Guha}}, \bibinfo {author} {\bibfnamefont {G.}~\bibnamefont
  {Dauphinais}},  \emph {et~al.},\ }\href@noop {} {\bibfield  {journal}
  {\bibinfo  {journal} {Quantum}\ }\textbf {\bibinfo {volume} {5}},\ \bibinfo
  {pages} {392} (\bibinfo {year} {2021})}\BibitemShut {NoStop}%
\bibitem [{\citenamefont {Fukui}\ \emph {et~al.}(2018)\citenamefont {Fukui},
  \citenamefont {Tomita}, \citenamefont {Okamoto},\ and\ \citenamefont
  {Fujii}}]{PhysRevX.8.021054}%
  \BibitemOpen
  \bibfield  {author} {\bibinfo {author} {\bibfnamefont {K.}~\bibnamefont
  {Fukui}}, \bibinfo {author} {\bibfnamefont {A.}~\bibnamefont {Tomita}},
  \bibinfo {author} {\bibfnamefont {A.}~\bibnamefont {Okamoto}}, \ and\
  \bibinfo {author} {\bibfnamefont {K.}~\bibnamefont {Fujii}},\ }\href
  {\doibase 10.1103/PhysRevX.8.021054} {\bibfield  {journal} {\bibinfo
  {journal} {Phys. Rev. X}\ }\textbf {\bibinfo {volume} {8}},\ \bibinfo {pages}
  {021054} (\bibinfo {year} {2018})}\BibitemShut {NoStop}%
\bibitem [{\citenamefont {Marandi}\ \emph {et~al.}(2014)\citenamefont
  {Marandi}, \citenamefont {Wang}, \citenamefont {Takata}, \citenamefont
  {Byer},\ and\ \citenamefont {Yamamoto}}]{marandi2014network}%
  \BibitemOpen
  \bibfield  {author} {\bibinfo {author} {\bibfnamefont {A.}~\bibnamefont
  {Marandi}}, \bibinfo {author} {\bibfnamefont {Z.}~\bibnamefont {Wang}},
  \bibinfo {author} {\bibfnamefont {K.}~\bibnamefont {Takata}}, \bibinfo
  {author} {\bibfnamefont {R.~L.}\ \bibnamefont {Byer}}, \ and\ \bibinfo
  {author} {\bibfnamefont {Y.}~\bibnamefont {Yamamoto}},\ }\href@noop {}
  {\bibfield  {journal} {\bibinfo  {journal} {Nature Photonics}\ }\textbf
  {\bibinfo {volume} {8}},\ \bibinfo {pages} {937} (\bibinfo {year}
  {2014})}\BibitemShut {NoStop}%
\bibitem [{\citenamefont {Kizmann}\ \emph {et~al.}(2019)\citenamefont
  {Kizmann}, \citenamefont {Guedes}, \citenamefont {Seletskiy}, \citenamefont
  {Moskalenko}, \citenamefont {Leitenstorfer},\ and\ \citenamefont
  {Burkard}}]{kizmann2019subcycle}%
  \BibitemOpen
  \bibfield  {author} {\bibinfo {author} {\bibfnamefont {M.}~\bibnamefont
  {Kizmann}}, \bibinfo {author} {\bibfnamefont {T.~L. d.~M.}\ \bibnamefont
  {Guedes}}, \bibinfo {author} {\bibfnamefont {D.~V.}\ \bibnamefont
  {Seletskiy}}, \bibinfo {author} {\bibfnamefont {A.~S.}\ \bibnamefont
  {Moskalenko}}, \bibinfo {author} {\bibfnamefont {A.}~\bibnamefont
  {Leitenstorfer}}, \ and\ \bibinfo {author} {\bibfnamefont {G.}~\bibnamefont
  {Burkard}},\ }\href@noop {} {\bibfield  {journal} {\bibinfo  {journal}
  {Nature Physics}\ }\textbf {\bibinfo {volume} {15}},\ \bibinfo {pages} {960}
  (\bibinfo {year} {2019})}\BibitemShut {NoStop}%
\bibitem [{\citenamefont {Riek}\ \emph {et~al.}(2017)\citenamefont {Riek},
  \citenamefont {Sulzer}, \citenamefont {Seeger}, \citenamefont {Moskalenko},
  \citenamefont {Burkard}, \citenamefont {Seletskiy},\ and\ \citenamefont
  {Leitenstorfer}}]{riek2017subcycle}%
  \BibitemOpen
  \bibfield  {author} {\bibinfo {author} {\bibfnamefont {C.}~\bibnamefont
  {Riek}}, \bibinfo {author} {\bibfnamefont {P.}~\bibnamefont {Sulzer}},
  \bibinfo {author} {\bibfnamefont {M.}~\bibnamefont {Seeger}}, \bibinfo
  {author} {\bibfnamefont {A.~S.}\ \bibnamefont {Moskalenko}}, \bibinfo
  {author} {\bibfnamefont {G.}~\bibnamefont {Burkard}}, \bibinfo {author}
  {\bibfnamefont {D.~V.}\ \bibnamefont {Seletskiy}}, \ and\ \bibinfo {author}
  {\bibfnamefont {A.}~\bibnamefont {Leitenstorfer}},\ }\href@noop {} {\bibfield
   {journal} {\bibinfo  {journal} {Nature}\ }\textbf {\bibinfo {volume}
  {541}},\ \bibinfo {pages} {376} (\bibinfo {year} {2017})}\BibitemShut
  {NoStop}%
\bibitem [{\citenamefont {Kim}\ and\ \citenamefont
  {Kumar}(1994)}]{Prem_Kumar_squeezing}%
  \BibitemOpen
  \bibfield  {author} {\bibinfo {author} {\bibfnamefont {C.}~\bibnamefont
  {Kim}}\ and\ \bibinfo {author} {\bibfnamefont {P.}~\bibnamefont {Kumar}},\
  }\href@noop {} {\bibfield  {journal} {\bibinfo  {journal} {Physical review
  letters}\ }\textbf {\bibinfo {volume} {73}},\ \bibinfo {pages} {1605}
  (\bibinfo {year} {1994})}\BibitemShut {NoStop}%
\bibitem [{\citenamefont {Dall'Arno}\ \emph {et~al.}(2010)\citenamefont
  {Dall'Arno}, \citenamefont {D'Ariano},\ and\ \citenamefont
  {Sacchi}}]{PhysRevA.82.042315}%
  \BibitemOpen
  \bibfield  {author} {\bibinfo {author} {\bibfnamefont {M.}~\bibnamefont
  {Dall'Arno}}, \bibinfo {author} {\bibfnamefont {G.~M.}\ \bibnamefont
  {D'Ariano}}, \ and\ \bibinfo {author} {\bibfnamefont {M.~F.}\ \bibnamefont
  {Sacchi}},\ }\href {\doibase 10.1103/PhysRevA.82.042315} {\bibfield
  {journal} {\bibinfo  {journal} {Phys. Rev. A}\ }\textbf {\bibinfo {volume}
  {82}},\ \bibinfo {pages} {042315} (\bibinfo {year} {2010})}\BibitemShut
  {NoStop}%
\bibitem [{\citenamefont {Fukui}\ \emph {et~al.}(2021)\citenamefont {Fukui},
  \citenamefont {Alexander},\ and\ \citenamefont {van
  Loock}}]{PhysRevResearch.3.033118}%
  \BibitemOpen
  \bibfield  {author} {\bibinfo {author} {\bibfnamefont {K.}~\bibnamefont
  {Fukui}}, \bibinfo {author} {\bibfnamefont {R.~N.}\ \bibnamefont
  {Alexander}}, \ and\ \bibinfo {author} {\bibfnamefont {P.}~\bibnamefont {van
  Loock}},\ }\href {\doibase 10.1103/PhysRevResearch.3.033118} {\bibfield
  {journal} {\bibinfo  {journal} {Phys. Rev. Research}\ }\textbf {\bibinfo
  {volume} {3}},\ \bibinfo {pages} {033118} (\bibinfo {year}
  {2021})}\BibitemShut {NoStop}%
\bibitem [{\citenamefont {Ralph}(1999)}]{ralph1999all}%
  \BibitemOpen
  \bibfield  {author} {\bibinfo {author} {\bibfnamefont {T.~C.}\ \bibnamefont
  {Ralph}},\ }\href@noop {} {\bibfield  {journal} {\bibinfo  {journal} {Optics
  letters}\ }\textbf {\bibinfo {volume} {24}},\ \bibinfo {pages} {348}
  (\bibinfo {year} {1999})}\BibitemShut {NoStop}%
\end{thebibliography}%

\end{document}